\documentclass{aa}  

\usepackage{graphicx}
\usepackage{txfonts}
\usepackage{lipsum}
\usepackage{subcaption}                                    
\usepackage{lscape}                                          
\usepackage{placeins}
\usepackage{hyperref}
\usepackage[dvipsnames]{xcolor}
\usepackage[normalem]{ulem}

\begin{document}

\title{Contrastive learning of extragalactic stellar streams}

\subtitle{Sculpting a latent space of representations with DES DR2 photometry}

\author{Ernesto Benitez-Walz\inst{1}\fnmsep\thanks{Corresponding author: benitez@astro.rug.nl}
    \and Jelle Mes\inst{2}
    \and Juan Miró-Carretero\inst{3,2}
    \and Koen Kuijken\inst{2}
    \and Amina Helmi\inst{1}
    }

\institute{Kapteyn Astronomical Institute, University of Groningen, Landleven 12, 9747 AD Groningen, The Netherlands
\and Leiden Observatory, Leiden University, P.O. Box 9513, 2300 RA Leiden, The Netherlands
\and Departamento de Física de la Tierra y Astrofísica, Universidad Complutense de Madrid, Plaza de las Ciencias 2, E-28040 Madrid, Spain}

\date{Received September 30, 20XX}
 
\abstract{We present a self-supervised approach for characterizing low-surface-brightness tidal features in wide-field imaging data by applying the nearest-neighbor contrastive learning of visual representations (NNCLR) algorithm to a curated subset of the Dark Energy Survey Data Release 2 (DES DR2). We constructed 38,334 cutouts of well-resolved galaxies in the $g,\,r,\,i$ bands, applying a novel ``tiered sigmoid scaling function'' to dynamically adjust image contrast according to the object’s signal-to-noise (S/N) and background levels. A supplemental labeled sample of 366 galaxies enabled the qualitative assessment of the learned embeddings. We trained a convolutional neural network with image augmentations including injection of simulated background stars and projected the resulting 512-dimensional representations into two dimensions using uniform manifold approximation and projection (UMAP) and its local density preserving variant (densMAP). We find that the NNCLR latent space recovers global trends corresponding to major merger features, yet does not reliably separate stellar streams without further supervision. To interpret the network’s implicit attention, we computed gradient-based saliency maps averaged over the full dataset: these reveal that the tiered sigmoid scaling effectively attenuates information from the center of the image cutouts, thereby suppressing the learning of high-surface-brightness features of each image cutout's central galaxy. Our study provides a blueprint for leveraging contrastive methods to mine forthcoming survey data for faint tidal substructure and highlights key preprocessing and interpretability considerations for robust stream detection.}


\maketitle
\nolinenumbers

\section{Introduction}
The formation and evolution of galaxies remain part of the most fundamental open questions in our understanding of large-scale structure. Given the dynamical nature of a galaxy’s morphology over the course of its formation, the resulting substructures left from this period trace a galaxy’s history \citep{barnes_encounters_1988, barnes_transformations_1992}. Stellar streams are the tidal remnants formed by gravitational interactions between a host galaxy and a satellite dwarf galaxy or globular cluster \citep{lynden-bell_ghostly_1995, johnston_disruption_1995, johnston_fossil_1996, Majewski1996, Helmi1999, helmi_stellar_2008, Stewart2008}. Such substructures can yield insights into the dark-matter component of galactic halos \citep{merrifield_measuring_1998, Koposov2010, ebrova_quadruple-peaked_2012, sanderson_analytical_2013, bonaca_milky_2014, Bowden2015, pearson_tidal_2015, Bovy2016, pearson_gaps_2017, bonaca_information_2018, pearson_mapping_2022, nibauer_constraining_2023, yavetz_stream_2023, nibauer_galactic_2025}.

The main challenge in observing extragalactic minor merger remnants is their low surface brightness relative to their host galaxy. Unlike the halo of the Milky Way, individual stars cannot be resolved for distant galaxies, and therefore deep imaging is required. Various extragalactic stellar streams have been identified by eye \citep{Atkinson2013, Sola2022, martinez-delgado_hidden_2023, Miro-Carretero2024, Martinez-Delgado2025}. However, visual inspection by humans alone is fundamentally impractical for analyzing the enormous volume of data produced by contemporary wide-field surveys such as the Kilo-Degree Survey \citep{de_jong_kilo-degree_2013}, the Hyper Supreme-Cam SSP Survey \citep{aihara_hyper_2018}, the Dark Energy Survey \citep{dark_energy_survey_collaboration_dark_2016}, the Ultraviolet Near-Infrared Optical Northern Survey \citep{gwyn_unions_2025}, the Legacy Survey of Space and Time \citep{Ivezic2019}, and the Euclid Wide Survey \citep{EuclidWideSurvey}, which collectively contain hundreds of millions of sources across thousands of square degrees.

Previous attempts at leveraging modern supervised methods of deep learning have obtained mixed results in the detection of faint stellar streams from minor mergers, largely due to the difficulty of the machine vision task and the scarcity of labeled training data. Simulated data often introduce biases that lead to low reliability when applied to real observational data \citep{Sanchez2023}. As such, citizen science efforts such as Galaxy Zoo \citep{Lintott2008} played a significant role in expanding the quantity of human-inspected image cutouts from large-scale surveys; however, streams pose a more difficult challenge than galaxies regarding detection and classification. Nonetheless, \cite{walmsley_identification_2019} achieved a binary classification with CFHTLS-Wide Survey \citep{gwyn_canadafrancehawaii_2012} image cutouts, and \cite{Gordon2024} recently demonstrated a method of supervised multiclass classification of low-surface-brightness features using the classifications from a visual inspection campaign of DECaLS \citep{Dey2019} photometric data. 

Contrastive learning is a self-supervised method of machine learning in which large unlabeled datasets are used to pretrain neural-network encoders to provide a ``latent space of representations'' in which similar features are mapped close together. When used in conjunction with labeled datasets in downstream classification tasks, these methods have been successfully applied to detect the tidal remains of major mergers, outperforming supervised deep neural networks in the low-label-count regime \citep[see][]{Desmons2024}. We aim to study the effect of several hyperparameter choices designed to steer the neural network's implicit attention toward extracting features that are useful in the detection of stellar streams.

We investigated the resulting latent space of the algorithm known as nearest-neighbor contrastive learning of visual representations (NNCLR) when trained on a subset of photometric data from the second data release of the Dark Energy Survey \citep{DES_DR2} (DES DR2). We used a collection of image cutouts containing known stellar streams from \cite{Miro-Carretero2024}, along with physical parameters of each host galaxy from the DES DR2 and Galaxy Zoo DECaLS \citep{Walmsley2021} catalogs to guide our inspection of the resulting embeddings. 

In Section \ref{sec:methodology}, we outline the subset of the DES DR2 we gathered and preprocessed, along with a novel scaling to enhance low-surface-brightness features. We also fully describe the contrastive learning pipeline we used to create and interpret the lower dimensional embeddings. Section \ref{sec:results} presents the model output gradients used in the interpretation of the model's implicit attention, along with the resulting latent space of representations when projected onto two dimensions. We also discuss some model limitations and our use of image queries, highlighting the importance of scaling-function selection when applying contrastive learning in the search for low-surface-brightness features. 

\section{Methodology}\label{sec:methodology}
\subsection{Data query}\label{subsec:des_dr2_subset}
The Dark Energy Survey is a ground based, six year long observational campaign by the Dark Energy Camera at the Cerro Tololo Inter-American Observatory in Chile. The $\sim5000\text{ deg}^2$ footprint of the southern galactic cap in five photometric bands ($g,\, r,\, i,\, z,\, Y$) provides ample targets, with a median imaging depth of 24.7, 24.4, and 23.8 mag in the $g,\,r,\,i$ bands at a signal-to-noise ratio (S/N) of 10 in a 1.95" aperture \citep{DES_DR2}.

The \texttt{des\_dr2.main} table contains 543 million galaxies and 145 million stars; we needed to reduce this set to ensure that each object would be a well-resolved galaxy. To this end, we filtered the dataset to uniquely include galaxies by using the \texttt{extended\_coadd\_class} parameter outlined in \cite{DES_DR2}. In essence, this parameter is a sum of boolean statements formed by various SExtractor \citep{Bertin1996} parameters that classify sources into four bins representing confidence levels of the object's morphology. We set a requirement of $\texttt{extended\_coadd\_class} > 2.5$, encompassing the "likely galaxies" and "high-confidence galaxies" bins. We further reduced the dataset by applying the following selection criteria:

\begin{itemize}
    \item $\texttt{flux\_radius\_r} > 3''$
    \item $13 < \texttt{mag\_auto\_\{g,r,i\}} < 17$
    \item $\texttt{mag\_auto\_r\_dered} < 18$
    \item $\texttt{imaflags\_iso\_\{g,r,i\}} = 0$
    \item $\texttt{flags\_\{g,r,i\}} < 4$,
\end{itemize}

\noindent where $\texttt{flux\_radius\_r}$ is the $r$-band half-light radius; $\texttt{mag\_auto\_\{g,r,i\}}$ is the SExtractor aperture magnitude in the $g,\,r,\,i$ bands for an elliptical aperture based on the Kron radius of the object; $\texttt{mag\_auto\_r\_dered}$ is the aperture magnitude in the $r$-band with the same aperture, dereddened via the prescription found in \cite{Schlegel1998}; \texttt{imaflags\_iso\_\{g,r,i\}} is a flag for missing and/or flagged pixels in single-epoch $g-,\,r-,\, \text{ and } i-$band images; and $\texttt{flags\_\{g,r,i\}}$ is a cumulative integer flag for boolean SExtractor flags such that values below four are well behaved. With the given selection criteria, we obtained a subset of the DES DR2 catalog consisting of 45,688 galaxies before proceeding with further quality cuts as outlined below. 

\subsection{Image cutout fetching and preprocessing}\label{subsec:image_preprocessing}
For each galaxy in our DES DR2 subset, we created a cutout centered at the right ascension and declination of the galaxy in the $g,\, r, \text{ and }\,i$ bands. The cutouts were obtained from the coadded image tiles of the DES DR2, where we selected the images with the largest exposure time to maximize imaging depth. We scaled the size of the cutout by 40 times the $r$-band half-light radius; the constant multiplier was heuristically determined to have the central object span one-third of the image. This procedure ensures enough of the field of view remains apparent to allow for the detection of surrounding low-surface-brightness features. The resulting distribution of image cutouts has a median area of $\sim680$ pixels$^2$ ($\sim3'$) and follows a power law, with a minimum and maximum image size of $\sim450$ pixels$^2$ ($\sim2'$) and $\sim4200$ pixels$^2$ ($\sim18'$), respectively. Each image was then downsampled to $256^2$ pixels via bilinear interpolation, yielding an input space with a dimensionality of $3\times256^2$ when accounting for the color information. The usage of different color bands introduces a ``speckling'' effect \citep{willett_galaxy_2017} in the background due to the high per-channel variance. To attenuate this effect, we used the same prescription as \cite{Walmsley2021}, where pixels with high per-channel variance are suppressed.

As we aim for the neural-network encoder to maximize the extraction of information from the surrounding near field of the central galaxy in each cutout, we scaled the image cutouts with a novel tiered sigmoid scaling function. The tiered sigmoid scaling function is a modified logistic function, defined by
\begin{equation}
    f(x_{k,b})=\frac{1}{1+\exp(-\Omega_{k,b})}, \label{eqn:2.1:tiered_sigmoid_scaling}
\end{equation}
where $x_{k,b}$ is the $k^{\text{th}}$ image in the $b\in\{g,\,r,\,i\}$ band. The $\Omega_{k,b}$ parameter is given by
\begin{align}
    \Omega_{k,b}&\equiv\frac{x_{k,b}-\mu_{j,b}\rho_{j,k,b}}{\sigma_{j,b}\rho_{j,k,b}}, \label{eqn:2.2:omega_param} \\
    \rho_{j,k,b}&\equiv\frac{\text{S/N}_{j,b}}{\text{S/N}_{k,b}},
\end{align}
where the $\mu$ and $\sigma$ hyperparameters are the center and growth rate of the logistic function. The S/N of the central galaxies are calculated from the \texttt{des\_dr2.main} table by using the SExtractor automated aperture photometry\footnote{\url{https://sextractor.readthedocs.io/en/des_dr1/Photom.html}} measurement for a given band \texttt{flux\_auto\_\{g,r,i\}} and dividing by the error of this measurement, \texttt{fluxerr\_auto\_\{g,r,i\}}.

The $j$ subscript in Equation \ref{eqn:2.2:omega_param} defines the different tiers in the tiered sigmoid scaling function. We split our dataset into four magnitude bins per band. Here, we referred to the \texttt{des\_dr2.main} table's \texttt{mag\_auto\_\{g,r,i\}} parameter for the magnitude of each image cutout's central galaxy. \texttt{mag\_auto\_\{g,r,i\}} is the magnitude obtained via the same elliptical model used by \cite{DES_DR2} to measure \texttt{flux\_auto\_\{g,r,i\}} and \texttt{fluxerr\_auto\_\{g,r,i\}}. Each bin spans one magnitude; it is these magnitude bins that we refer to as the different tiers of the tiered sigmoid scaling function. Our choice of binning stems from the varying local sky background levels of the image cutouts yielding no single pair of $\mu,\,\sigma$ values that provide an ideal dynamic range for observing stellar streams in every image cutout. We calibrated the center and growth rate of the tiered sigmoid scaling function per tier and per band by gathering photometric statistics from elliptical apertures encompassing stellar streams reported by \cite{Miro-Carretero2024}.

In addition to the DES DR2 catalog subset, we gathered a subset of the catalog of galaxies inspected by \cite{Miro-Carretero2024}, including galaxies determined not to host stellar streams. Due to limitations of our dataset gathering pipeline for neural network input, of the total 689 galaxies inspected by \cite{Miro-Carretero2024}, we ingested 528 cutouts for which the requirements of square cutouts with one-third central object scaling are met. Of these 528 objects, 47 contain stellar streams. We referred to the 47 objects with stellar streams as ``positives,'' while the remaining are referred to as ``negatives,'' forming the labeled dataset. 

For each tier of the tiered sigmoid scaling function, we randomly selected up to three galaxies from the set of positives for each band. We chose to randomly sample the set of positives to decrease the bias toward any tidal debris morphology; our choice of galaxies is outlined in Table \ref{tab:A1:sigmoid_host_galaxies} of Appendix \ref{appendix:calibrating_the_tiers}. We performed a visual inspection of the $r$-band image to choose an elliptical aperture surrounding the stellar stream in each sampled positive image cutout. We then used this elliptical aperture on all three bands to gather median and median absolute deviation statistics of the pixel values (in picomaggies)\footnote{\url{https://www.sdss3.org/dr8/algorithms/magnitudes.php}} within the aperture, which we averaged per tier and report in Table \ref{tab:A1:sigmoid_statistics} of Appendix \ref{appendix:calibrating_the_tiers}. A drawback of the tiers of our scaling method is the introduction of a category that the network can discern (see Appendix \ref{appendix:tier_randomization_ablation}). We therefore introduced a 30\% probability of randomly selecting a tier per image. Through this method, we tuned our image scaling hyperparameters to yield the best dynamic range for streams similar to those in the \cite{Miro-Carretero2024} sample. 

\begin{figure}
    \centering
    \includegraphics[width=\columnwidth]{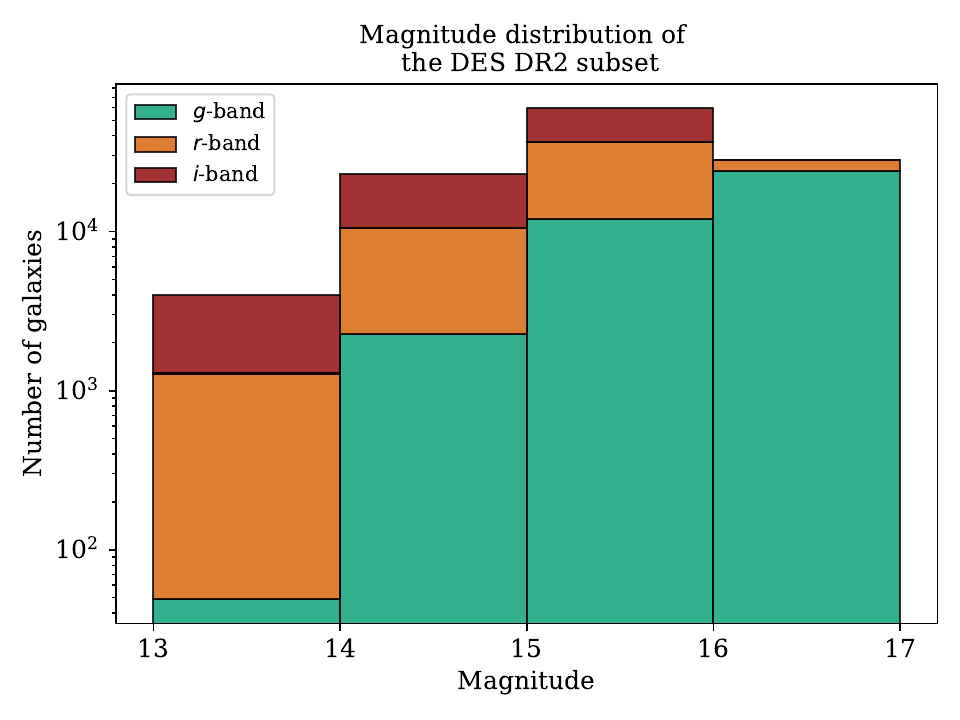}
    \caption{Histogram of objects per tiered sigmoid scaling function bin, as validated by the procedure described in Section \ref{subsec:image_preprocessing}. The  bins represent the DES DR2 subset, with a total of 38,334 unlabeled objects used for training.}
    \label{fig:2.1:sigmoid_tiers}
\end{figure}

Finally, we performed a quality cut and removed cutouts in which bright stars dominated the background. As discussed in \cite{DES_DR1}, on a few occasions the cutouts were missing data in one of the three bands such that a portion of the image was filled with null values. Despite the \texttt{imaflags\_iso\_\{g,r,i\}} parameter filtering cutouts for which there is missing data, we performed a sanity check to ensure that the central object was present in all three bands by requiring the central 10x10 pixels to contain no null values. We then required that after applying the tiered sigmoid scaling function, none of the color channels surpassed an average value of 0.4, removing most images with a bright object in the field. Finally, we ensured no \texttt{NaN} or \texttt{inf} values were present. After applying these cuts to the DES DR2 subset and the labeled dataset, we were left with 38,334 unlabeled and 366 labeled valid objects, respectively. The validated images were distributed into the tiers of the tiered sigmoid scaling function as shown in Figure \ref{fig:2.1:sigmoid_tiers}.

\begin{figure}
    \centering
    \includegraphics[width=\linewidth]{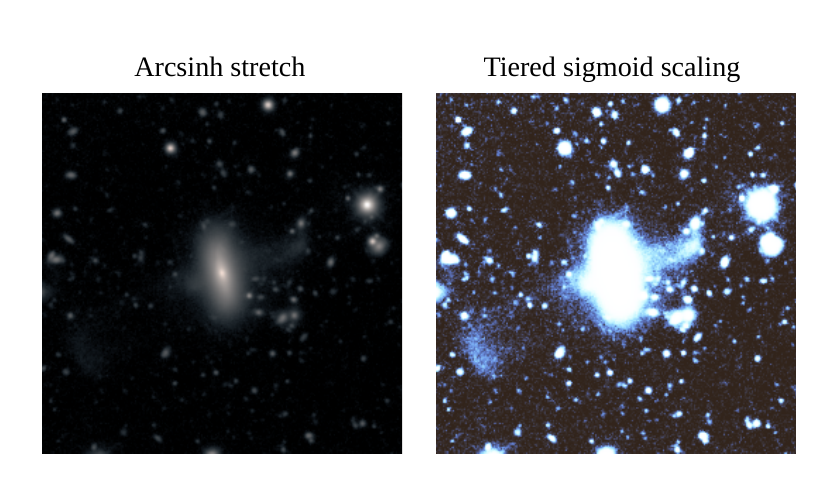}
    \caption{From visual inspection of our dataset, the galaxy LEDA 131334 contains a candidate stellar stream shown above with the (left) arcsinh stretch used by \cite{Desmons2024} and (right) tiered sigmoid scaling function.}
    \label{fig:scaling_example}
\end{figure}

As a point of comparison between traditional scaling functions and the tiered sigmoid scaling function we chose the arcsinh stretch applied by \cite{Desmons2024}, since \cite{Desmons2024} used the arcsinh stretch in the search for extragalactic tidal features using contrastive learning. The arcsinh stretch is defined by
\begin{equation}
    f(x_{b}) = \sinh^{-1}\bigg(\frac{x_b}{3\sigma_b}\bigg),
\end{equation}
where $\sigma_b$ is the standard deviation of band $b$, obtained via the median absolute deviation of 1000 randomly sampled image cutouts. Figure \ref{fig:scaling_example} demonstrates an example image scaled with the arcsinh stretch and our tiered sigmoid scaling function. 

\subsection{Nearest-neighbor contrastive learning of visual representations}\label{subsec:nnclr}
\begin{figure*}[h!]
    \centering
    \includegraphics[width=\textwidth]{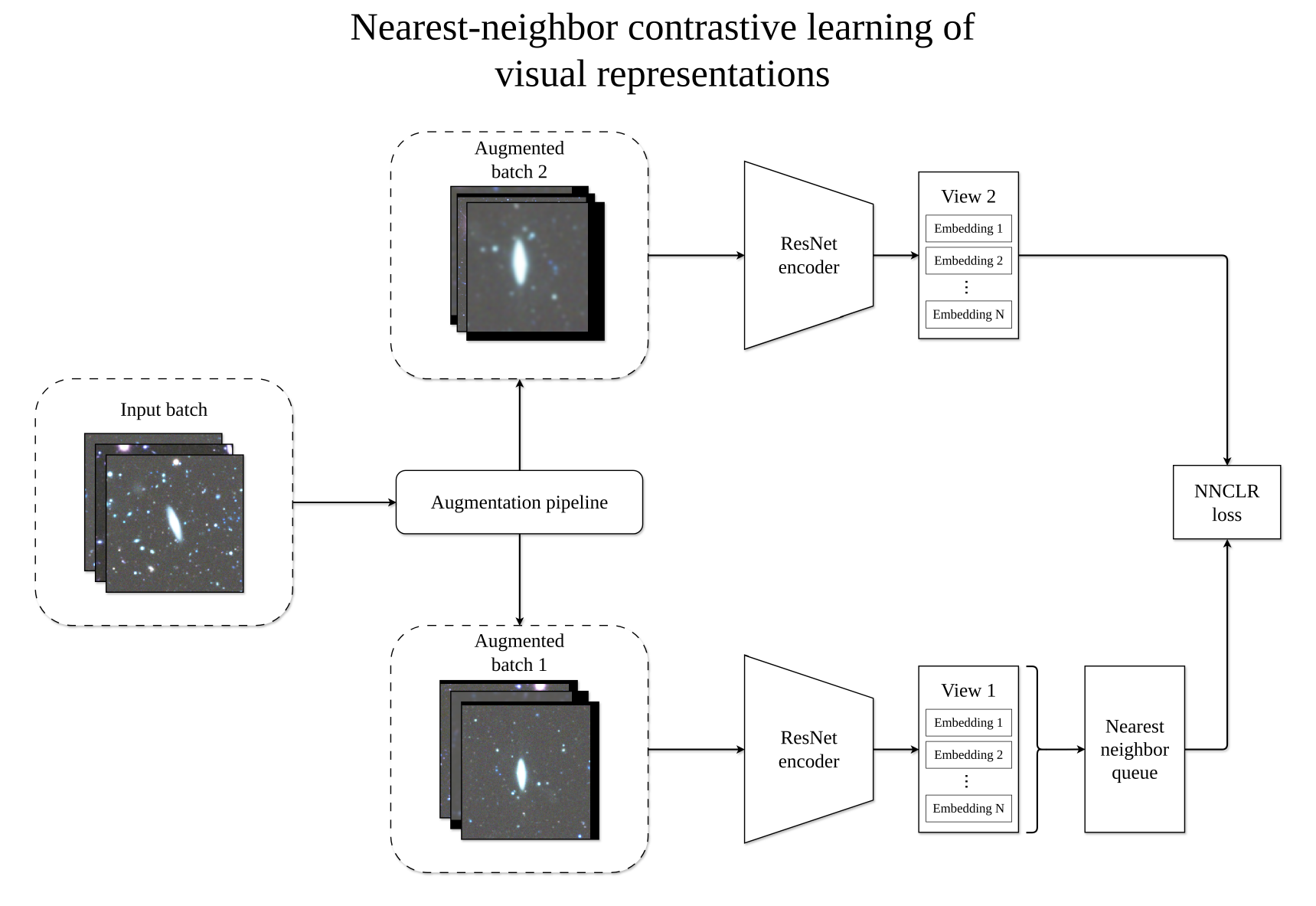}
    \caption{Schematic of NNCLR algorithm as outlined by \cite{Dwibedi2021}. A batch of images was passed through the augmentation pipeline described in Section \ref{subsubsec:augmentations}, producing two augmented versions of each input image. Each augmented batch is then passed through the NNCLR encoder, where we chose the residual neural-network architecture (ResNet-18 \citep{Kaiming2016}; see Section \ref{subsec:nnclr}). The resulting embedding vectors, referred to as ``views,'' are $\ell_2$ normalized (Euclidean norm) and compared to the nearest neighbor of a queue formed by the embedding vectors of other images in the dataset. The NNCLR loss function quantifies the similarity between the views and the nearest neighbor via cosine similarity, and it is used by the stochastic gradient descent optimizer to update the model weights accordingly.}
    \label{fig:2:NNCLR}
\end{figure*}
Nearest-neighbor contrastive learning of visual representations (NNCLR) is a contrastive learning algorithm which uses the nearest neighbor operation to improve the generalization of the embedding vectors within the latent space of the encoder's output in representing different views of objects closely related in semantics \citep{Dwibedi2021}. A full review of the algorithm along with its response to ablations is provided in \cite{Dwibedi2021}; here, we summarize the procedure as it was implemented in our data processing pipeline. 

The NNCLR algorithm is based on ``instance discrimination,'' where pairs of a single image are created by passing the image through a set of random transformations (known as ``image augmentations''). These transformations are selected to render the NNCLR encoder invariant to the chosen transformations; further discussion on the choice of transformations included in our augmentation pipeline is given in Sections \ref{subsubsec:augmentations} and \ref{subsubsec:starsim}. Each image in the image pair is then individually passed through the NNCLR encoder and a projection head, outputting an embedding vector referred to as a ``view.'' 

Figure \ref{fig:2:NNCLR} presents an overview of the steps taken by the NNCLR algorithm. Instead of directly comparing the two views of the same image via a contrastive loss function, NNCLR compares one of the two views with the nearest neighbor of a set of views from other image pairs in the dataset. This set of different views is generated as a queue of views from other image pairs on a first-in-first-out basis; upon finishing the loss computation and updating the model weights, one of the two views is added to the queue and the oldest view is removed. The loss function is therefore defined by 
\begin{equation}
    \mathcal{L}_i^{\text{NNCLR}}=-\log\frac{\exp(\text{NN}(z_i,Q)\cdot p_i^+/\tau)}{\sum_{k=1}^n\exp(\text{NN}(z_i,Q)\cdot p_k^+/\tau)}
,\end{equation}
where $z_i$ represents the first embedding vector (View 1 in Figure \ref{fig:2:NNCLR}) of the $i^{\text{th}}$ image pair in a batch of $n$ image pairs, $p_i^+$ is the second embedding vector that is additionally passed through a prediction head (View 2 in Figure \ref{fig:2:NNCLR}), Q is the queue of embeddings, NN represents the nearest-neighbor operation, and $\tau$ is a softening parameter for the exponential (here we set $\tau=0.1$).

We chose the PyTorch Lightning\footnote{\url{https://lightning.ai/docs/pytorch/stable/}} framework to implement the deep-learning portions of this study. Our choice of NNCLR encoder for reducing the dimensionality of each augmented image pair was an 18-layer deep residual neural network (a ResNet-18) \citep{Kaiming2016}. Residual neural networks employ blocks consisting of several nonlinear layers before dimensionality reduction and skip connections, reducing the risk of vanishing and/or exploding gradients with increasing network depth. The projection head consisted of three fully connected layers with [512, 512, 128] nodes, with the first and second layers having ReLU activation and all three layers having batch normalization. The prediction head had two fully connected layers of size [2048, 128], of which only the first layer had ReLU activation and batch normalization. Throughout testing, we used a batch size of 128 images and performed training with an Nvidia A100 40GB GPU with 16-bit automatic mixed precision enabled.

We optimized the neural networks with \texttt{AdamW} \citep{Loshchilov2017}, which is a variant of the popular \texttt{Adam} adaptive stochastic gradient descent algorithm with an improved weight-decay implementation. The learning rate was initialized at $10^{-3}$ and cosine annealed to $10^{-4}$ over 50 epochs of training. Furthermore, because we aimed to compare the effects of different scaling functions on the resulting latent space of representations, we applied a global seed for all \texttt{numpy}, \texttt{pytorch}, and \texttt{python.random} objects and utilized the same initial model parameters to reduce stochasticity throughout optimization and downstream analysis between scaling functions.

\subsubsection{Augmentations}\label{subsubsec:augmentations}
As mentioned, NNCLR belongs to the instance-discrimination category of contrastive learning and therefore requires different representations of the same semantic class to learn which features in the dataset's images are relevant. We therefore designed a pipeline that produces pairs of images with different random transformations. We selected the augmentations to render the NNCLR encoder invariant to the information content of the images that we do not wish the network to discriminate against. For the sake of direct comparison between the latent representations of the tiered sigmoid and arcsinh scaling functions, we applied the same randomly chosen augmentations to any given sample, ensuring that neither scaling function randomly receives a more beneficial set of augmented images.

The augmentation pipeline includes the following random transformations, listed in order of application:
\begin{itemize}
    \item Random star simulation and injection ($\mathcal{U}(10,\,50)$ stars)
    \item Noise injection, where the noise is sampled from $\mathcal{N}(0,N\sigma_\text{MAD})$ for each band, and where $N\sim\mathcal{U}(1,\,3)$ as per \cite{hayat_self-supervised_2021}.
    \item Random spatial translation with center cropping and resizing ($x_\text{shift},\,y_\text{shift}\sim\mathcal{U}(0,\,32),\mathcal{U}(0,\,32)$ pixels of translation, 30\% probability) 
    \item Random horizontal flipping (50\% probability)
    \item Random vertical flipping (50\% probability)
    \item Random rotation by $\theta\sim\mathcal{U}(-180^\circ, 180^\circ)$
    \item Random grayscaling (5\% probability).
\end{itemize}

We randomly generated stars to reduce the NNCLR encoder's sensitivity to background stars, as further described in Section \ref{subsubsec:starsim}. The noise injection augmentation serves to decrease the NNCLR encoder's sensitivity to varying background noise across different image cutouts; \cite{hayat_self-supervised_2021} found the noise injection to be the most relevant augmentation for improving the quality of self-supervised visual representations of astronomical images. In our DES DR2 subset, not every galaxy was completely centered within its respective image cutout; as such, we included the spatial translation augmentation with cropping and resizing to account for these outliers. Finally, we introduced a low probability of grayscaling (e.g., randomly selecting the $r$-band so that the color channels of an image went from $\{g,\,r,\,i\}$ to $\{r,\,r,\,r\}$) to reduce the network's sensitivity to color information. We wished for the NNCLR encoder to utilize color information---which is important for tasks such as distinguishing spiral arms from stellar streams---while balancing the spatial and morphological information of the images when creating the embedding vectors so that color does not dominate the way in which the embeddings are distributed in the latent space of representations.

\subsubsection{Star simulator}\label{subsubsec:starsim}
The star simulator augmentation creates realistic fake foreground stars at random positions in the image cutouts of our DES DR2 subset. The realism is of particular importance for the NNCLR encoder, as we must avoid introducing the learned behavior of discriminating between fake simulated stars and real stars. In order to create realistic stars, we calculated an effective point spread function (ePSF) on a per-band, per-image basis. The ePSF was calculated by source extracting stars in each band of each image using the \texttt{DAOStarFinder} class of Photutils \citep{Stetson1987, photutils}. We then used the \texttt{EPSFBuilder} of Photutils to create each ePSF.

\begin{figure}[h!]
    \centering
    \includegraphics[width=\linewidth]{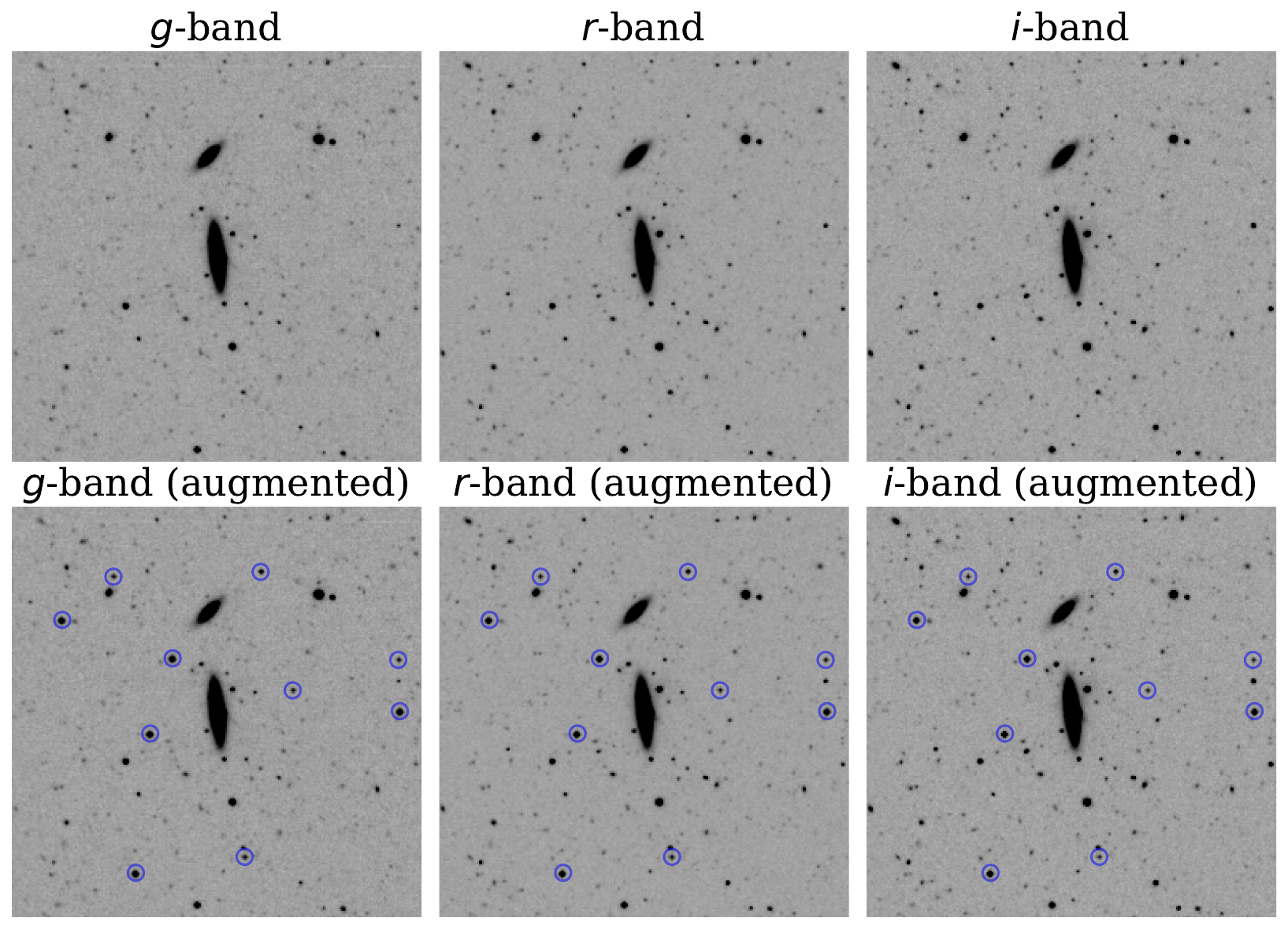}
    \caption{Randomly selected image (top row) prior to the star simulator augmentation and (bottom row) after applying the augmentation. For this example, we included blue apertures to ease identification of the generated fake stars; the apertures are not present for the images used to train the NNCLR models.}
    \label{fig:2:simulated_stars_example}
\end{figure}

We obtained realistic colors by sampling from the centroid of the apertures returned by the \texttt{DAOStarFinder} and adding a small uniform jitter to the color values. The color in each band was then used to initialize a $\mathcal{U}$(1, 3) pixel kernel, which was convolved with the ePSF of the image at each band. The last step was to sample random radial coordinates from the center of the image, where $\theta\sim\mathcal{U}(0,\,2\pi)$ and $r\sim\mathcal{U}(\frac{256}{3\times2},\,\frac{256}2{)}$ pixels, and add the fake star to the randomly chosen coordinates. The bounds of the radius were chosen to avoid generating the star within the central galaxy of the image cutout. Figure \ref{fig:2:simulated_stars_example} demonstrates a randomly chosen image cutout prior to and after augmenting the cutout with simulated stars.

\subsection{Saliency maps}\label{subsec:saliency_maps}
A common limitation of deep-learning models is their lack of interpretability. Because neural networks rely on numerous layers and a vast number of tunable parameters, it is often unclear how they process inputs to produce outputs. This opacity leaves us largely unaware of what internal operations the network performs during inference. However, recent advances have introduced models with more interpretable structures such as vision transformers \citep{Dosovitskiy2021}. These models include an explicit attention mechanism that allows the network to focus selectively on different parts of the input while also providing context about how those parts relate to the whole. This attention mechanism not only improves performance but also offers valuable insight into the model's decision-making process.

Although the residual neural networks used in this study do not include an explicit attention mechanism, we can still gain interpretability through implicit attention. Specifically, we investigated the behavior of the NNCLR encoder by computing a saliency map using the Jacobian derived from the backpropagation of gradients from the output of the NNCLR encoder, as described in \cite{Simonyan2014}. We chose this method of assessing saliency since it is robust for the visual interpretation of saliency, being both dependent on model parameters at various depths and (although irrelevant for unsupervised deep learning) randomized labels \citep{Adebayo2018}.  We therefore computed for each image cutout

\begin{equation}
    E_{\text{grad}}=\underset{g,r,i}{\text{max}}\Bigg(\Bigg|\frac{\partial ||S(x)||_{\ell_2}}{\partial x}\Bigg|\Bigg),
\end{equation}

\noindent where $||S(x)||_{\ell_2}$ is the Euclidean norm of the latent representation vector of image $x$. We used the maximum gradient values across the $g$, $r$, and $i$ color channels of the Jacobian to assess overall saliency. Finally, we scaled $E_{\text{grad}}$ to be between zero and unity before averaging $E_{\text{grad}}$ over the entire dataset. Unlike \cite{Simonyan2014}, in which only a single class score is used when calculating $E_\text{grad}$, the latent representation vector for any given image encodes many features of the image. The saliency map of a single image therefore cannot be used to determine if the model's implicit attention is focused on a tidal feature, since many other features will be highlighted as well (see Appendix \ref{appendix:single_image_saliency_maps}). Instead, this technique reveals which parts of the input most strongly influence the network’s output; a larger value of $E_\text{grad}$ for any given pixel represents the largest deviation from the model parameters. 

\subsection{Latent representation space projection}\label{subsec:umap}
The output of the NNCLR encoder is a high-dimensional feature vector. Because it is not possible to directly visualize data in such high dimensions, we reduced the embeddings to two dimensions for analysis. It is crucial that this dimensionality reduction preserves the structure and semantic relationships present in the original high-dimensional space. For this reason, we used the uniform manifold approximation and projection (UMAP) algorithm \citep{McInnes2018}, which is designed to preserve the topological structure of the data during projection.

Before interpreting the resulting UMAP plots from NNCLR embeddings, we briefly explain how UMAP works. The full method is detailed in \cite{McInnes2018}. UMAP treats the high-dimensional embeddings as samples from a continuous manifold. It models this manifold using sets of k-simplices (geometric shapes defined by k vertices). For example, a 0-simplex is a point (an embedding), and a 1-simplex is a line (a connection between two embeddings).

Using this framework, UMAP constructs a weighted graph where nodes represent embeddings and edges represent the proximity between them. Edge weights reflect the probability of connections, accounting for the fact that in high-dimensional space, distances between points become increasingly similar (a phenomenon known as the ``curse of dimensionality"). To handle this, UMAP makes the graph ``fuzzy" by modeling these connection probabilities rather than making hard connections. Finally, UMAP projects this fuzzy graph into two dimensions using a force-directed layout algorithm \citep{Kobourov2012}, which places nodes in such a way that edges have approximately equal length and nodes do not overlap. The result is a two-dimensional embedding that preserves the global and local structure of the original high-dimensional space as much as possible.

When interpreting a UMAP visualization, it is important to recognize that the values along the axes do not carry inherent semantic meaning. Instead, what can offer insight is the relative positioning of clusters, such that closer clusters may suggest similar features in the original embedding space. However, these spatial relationships can also be artifacts introduced by the projection process, and should therefore be interpreted with caution.

We also applied a variant of UMAP called densMAP \citep{narayan_assessing_2021} which uses a local density estimate of the high-dimensional input (the NNCLR embeddings in our case) as a regularization term when projecting to two dimensions. The resulting two-dimensional projection of densMAP therefore has a more accurate estimation of local density, and is beneficial when comparing the resulting latent space of representations formed by the NNCLR encoder trained on tiered-sigmoid-scaled images against the NNCLR encoder trained on arcsinh-scaled images.

\section{Results}\label{sec:results}
Here we discuss the features of the resulting latent spaces from having trained the NNCLR encoders on the dataset described in Sections \ref{subsec:des_dr2_subset} and \ref{subsec:image_preprocessing}. We inspect the saliency maps produced by the NNCLR encoders trained with the tiered-sigmoid-scaled and arcsinh-scaled images to determine which features each model learns from. Furthermore, we present various mean histograms of the Galaxy Zoo DECaLS \citep{Walmsley2021} \texttt{merging\_merger\_fraction} parameter with two-dimensional projections of the NNCLR embeddings from each encoder using UMAP and densMAP; moreover, we investigated the nearest neighbors of the latent representations for a galaxy with low-surface-brightness features.

\subsection{Saliency maps}
We first probed the features considered by the NNCLR encoder within the image cutouts using the gradients of the model output with respect to the input \citep{Simonyan2014, Adebayo2018}. Figure \ref{fig:3:saliency_maps_averaged} demonstrates the saliency maps obtained via the gradients of the model outputs of each model (trained separately on tiered-sigmoid-scaled and arcsinh-scaled images), averaged across the entire DES DR2 subset. As described in Section \ref{subsec:saliency_maps}, we observe the maximum gradients across the color channels of the input space to encapsulate the overall implicit attention of the models.

\begingroup
\renewcommand{\arraystretch}{2.0}
\begin{figure}[h!]
    \centering
    \begin{tabular}{c}
        \includegraphics[width=0.98\linewidth]{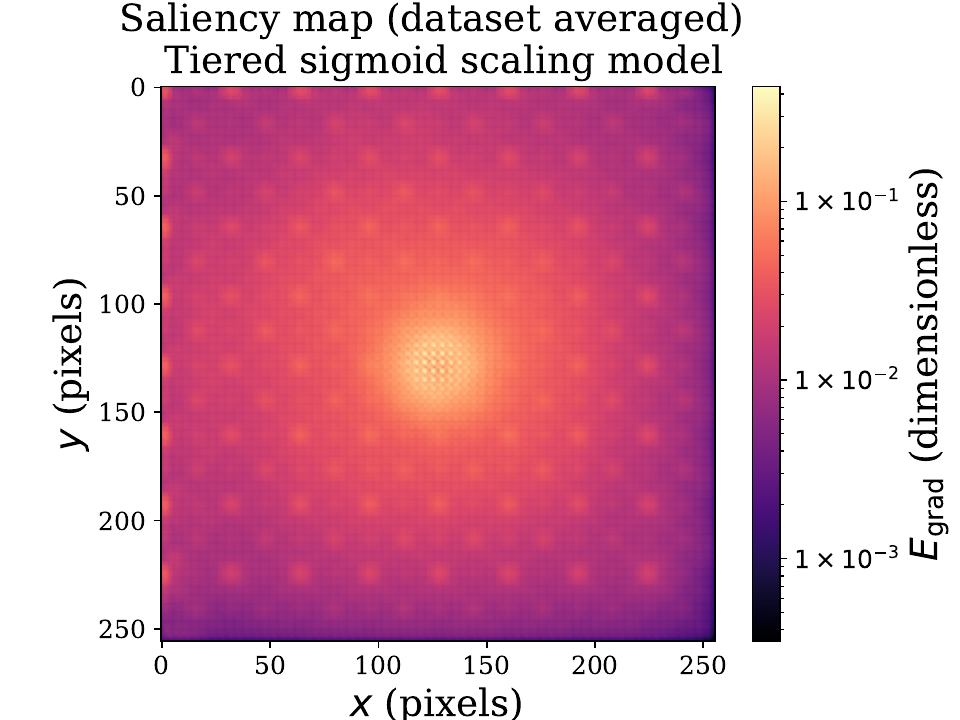} \\
        \includegraphics[width=0.98\linewidth]{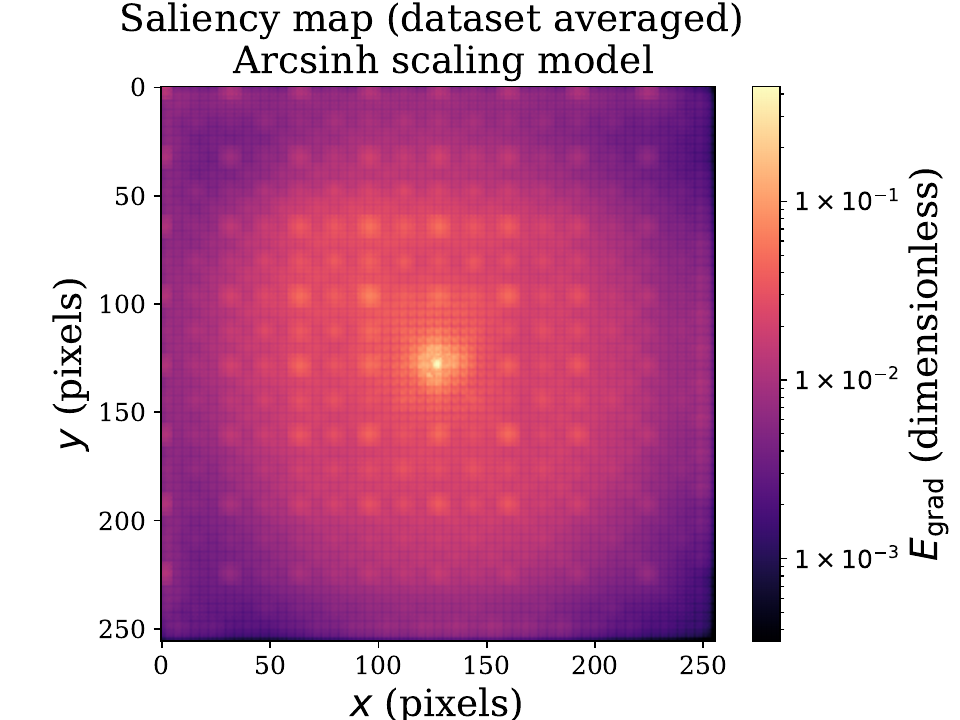}
    \end{tabular}
    \caption{Saliency map of model trained on the DES DR2 subset with the (top) tiered sigmoid scaling function and (bottom) the arcsinh stretch of \cite{Desmons2024}. These maps represent the gradients of the model output with respect to each input pixel. We used the maximum gradient values across color channels as per \cite{Simonyan2014} and averaged over the entire DES DR2 subset. The gradients were scaled to be between zero and unity. We observe the model trained with data that were tiered-sigmoid-scaled and most sensitive to changes within a broader area of the center, such that the model extracts less information from the central galaxy in each cutout. Note that the repeating patterns are artifacts due to the convolutional operations of the NNCLR encoder along with the skip connections of the residual neural-network architecture.}
    \label{fig:3:saliency_maps_averaged}
\end{figure}
\endgroup

The saliency maps in Figure \ref{fig:3:saliency_maps_averaged} show that the tiered sigmoid scaling function reduces a model’s ability to extract information from a broad region in the center of the image cutouts (approximately 100–150 pixels in both axes) compared with the model trained on the arcsinh-scaled dataset (approximately 115–135 pixels). This behavior arises from the inherent masking introduced by the tiered sigmoid scaling function; because each image cutout is explicitly scaled such that roughly one-third of the frame is occupied by the central galaxy (and the galaxy pixels are saturated at unity), any deviation from this value provides a strong learning signal. In contrast, the saliency map for the model trained on the arcsinh-scaled dataset exhibits high gradient values concentrated in a smaller central region of the image. As expected, arcsinh scaling preserves more information from the inner regions of the galaxies, except for the cores, which remain saturated due to their higher surface brightness.

\subsection{Latent space of representations}\label{subsec:umap_embeddings}
\begin{figure*}
    \centering
    \includegraphics[width=\textwidth]{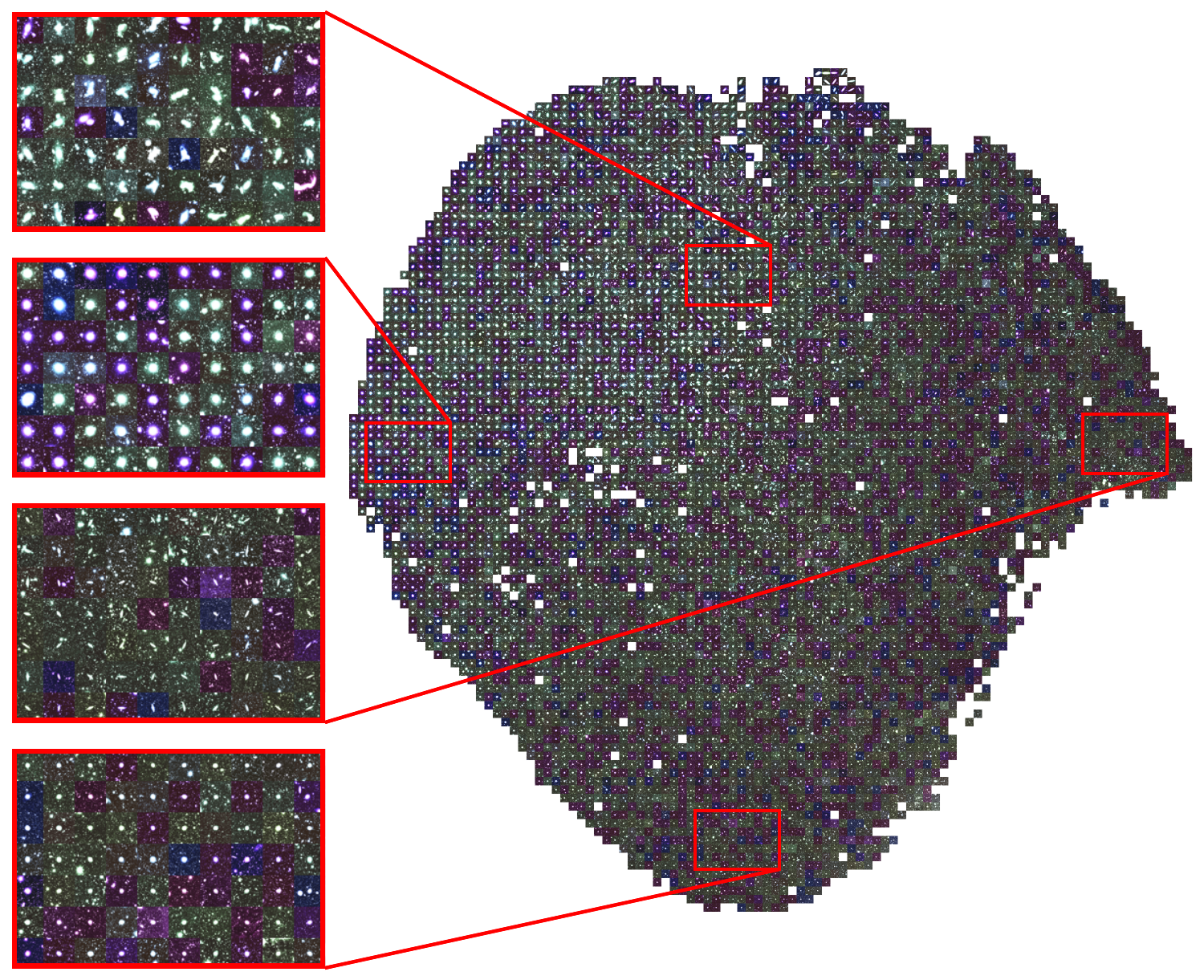}
    \caption{Mosaic of DES DR2 subset (described in Section \ref{subsec:des_dr2_subset} and \ref{subsec:image_preprocessing}) embedding vectors from the NNCLR encoder after being projected to two dimensions by UMAP. As in \cite{Desmons2024}, we binned the space into a $100\times100$ grid and randomly sampled a single object per bin to display. The zoomed-in view of boxes on the left-hand side show the distinct populations we can visually identify within the UMAP. From the top, the first box demonstrates galaxies that are visually identified to contain major merger features (and further confirmed by cross-matching with the \texttt{merging\_merger\_fraction} parameter from \cite{Walmsley2021}; see Figure \ref{fig:3:merging_merger_fraction_umap}). The second box contains galaxies that have a more extended light profile and transition more smoothly to the background as opposed to galaxies in the third and fourth boxes. Notably, the third and fourth boxes contain galaxies with similar features, but different ellipticities. It is difficult to teach a contrastive learning encoder to be invariant to ellipticity through unsupervised methods, as no image augmentation can recreate viewing the same galaxy from different perspectives. Note that the differences in overall color (red-green-blue) of the images are caused by the variance in the growth rate across color channels due to the tiers of the sigmoid scaling function (see Appendix \ref{appendix:tier_randomization_ablation}).}
    \label{fig:3:fiducial_mosaic}
\end{figure*}
The saliency maps of the previous subsection indicate that there is a substantial difference in the implicit attention between the models trained on tiered-sigmoid-scaled and arcsinh-scaled images. We therefore investigated the impact this effect has on the resulting latent representations using the UMAP and densMAP projections. 

We first studied the global structure of the latent space of representations by using the UMAP-projected NNCLR embeddings. Figure \ref{fig:3:fiducial_mosaic} presents a two-dimensional UMAP projection of the NNCLR embedding vectors of the model trained with tiered-sigmoid-scaled images, arranged into a mosaic of image cutouts as per \cite{hayat_self-supervised_2021}. By partitioning the projection plane into a $100\times100$ grid and randomly selecting one object per cell, coherent regions emerge in which galaxies with similar visual characteristics cluster together. We observe regions of the UMAP in which galaxies appear to be in the process of a major merger (first subpanel from top to bottom), which are in the transition region between galaxies that have a more extended light profile and are generally less elliptical (second versus third and fourth subpanels). 

\begin{figure*}
    \centering
    \begin{tabular}{cc}
        \includegraphics[width=\columnwidth]{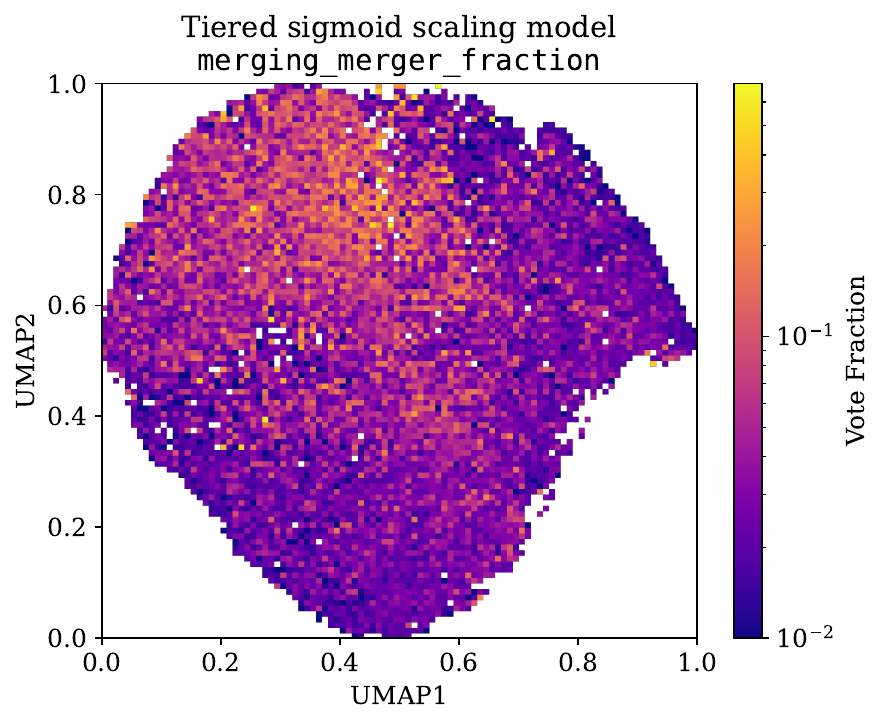} & 
        \includegraphics[width=\columnwidth]{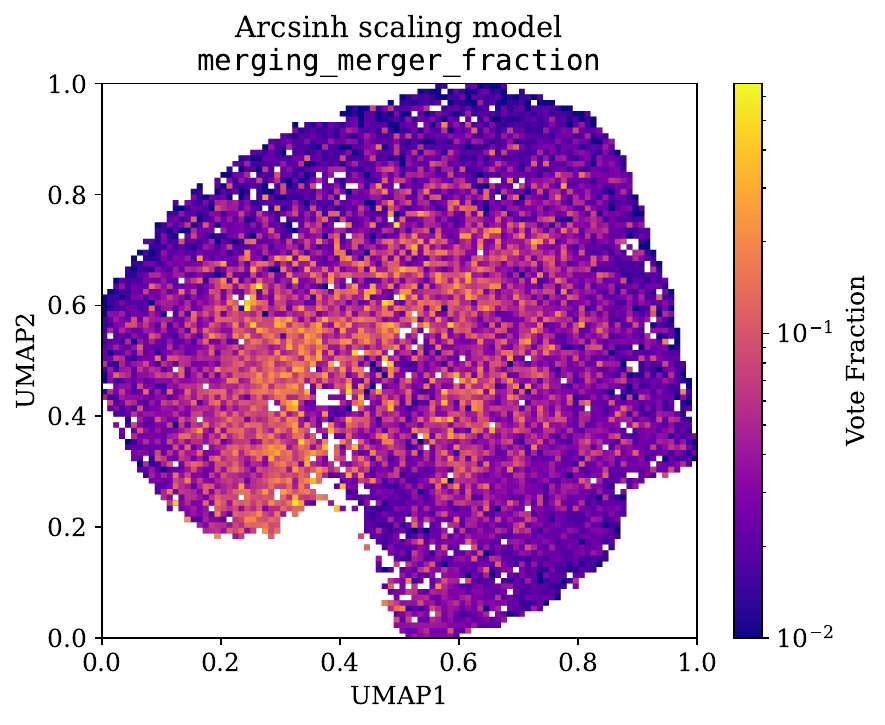}
    \end{tabular}
    \caption{Mean histograms of \texttt{merging\_merger\_fraction} Galaxy Zoo DECaLS parameter \citep{Walmsley2021} for the (left) UMAP projection of the tiered sigmoid scaling NNCLR model embeddings and (right) the UMAP projection of the NNCLR embeddings from the model trained on arcsinh-scaled images.}
    \label{fig:3:merging_merger_fraction_umap}
\end{figure*}

As a visual aid for this interpretation, we also examined UMAP projections binned by the Galaxy Zoo DECaLS \texttt{merging\_merger\_fraction} parameter \citep{Walmsley2021}. Figure \ref{fig:3:merging_merger_fraction_umap} demonstrates the increase in fractional votes for major merger features in the same region of the UMAP as identified in the top subpanel of Figure \ref{fig:3:fiducial_mosaic}. These projections are used purely as a qualitative diagnostic to highlight how known morphological trends map onto the learned embedding and to visually identify regions associated with distinct physical processes, such as mergers. Because standard UMAP prioritizes neighborhood structure and visual separability over faithful density preservation, we do not use these projections for quantitative inference.

\begin{figure*}
    \centering
    \begin{tabular}{cc}
        \includegraphics[width=\columnwidth]{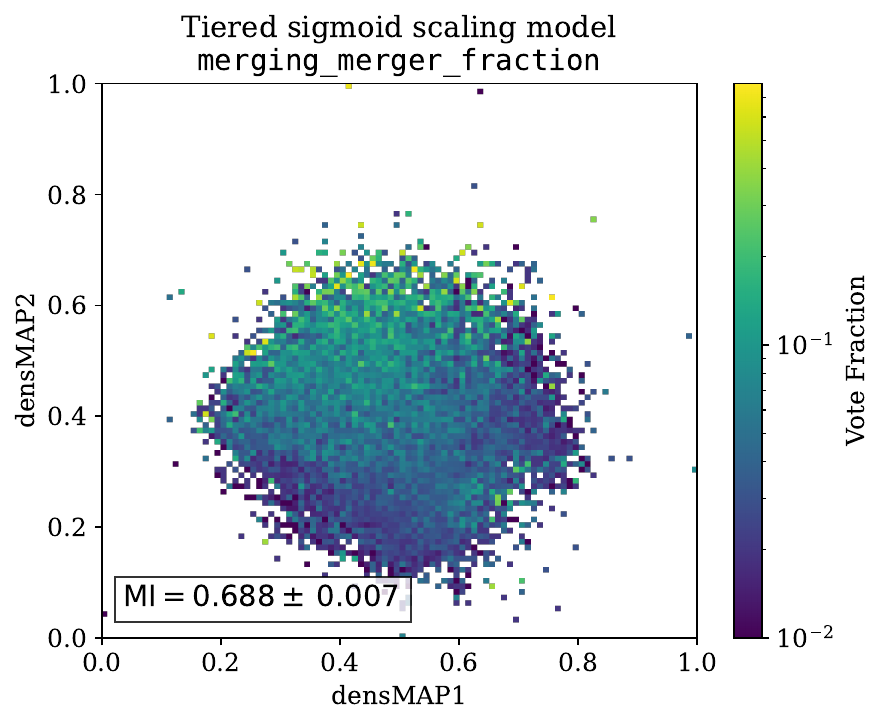} &
        \includegraphics[width=\columnwidth]{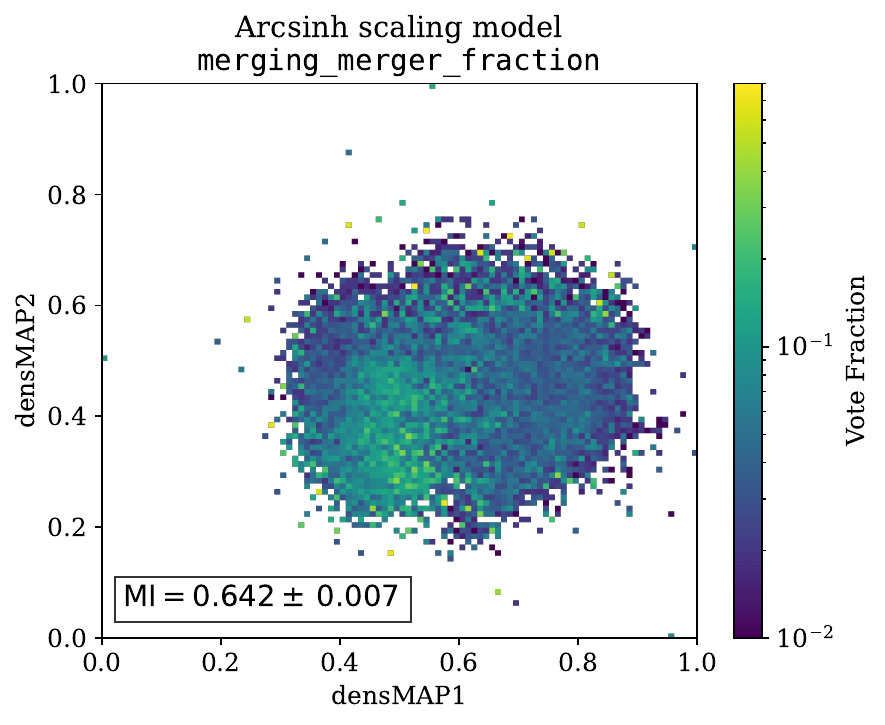}
    \end{tabular}
    \caption{Mean histograms of \texttt{merging\_merger\_fraction} Galaxy Zoo DECaLS parameter \citep{Walmsley2021} for the (left) densMAP projection of the tiered sigmoid scaling NNCLR model embeddings and (right) the densMAP projection of the NNCLR embeddings from the model trained on arcsinh-scaled images. The text box at the bottom left of both plots states the mutual information of the given histogram, as defined in Section \ref{subsec:umap_embeddings}.}
    \label{fig:3:merging_merger_fraction_densmap}
\end{figure*}

To quantify the degree to which there is clustering among the different regions in the projection, we computed the mutual information of the mean histogram formed by the \texttt{merging\_merger\_fraction} parameter when projected to two dimensions. For this purpose, we adopted densMAP rather than the standard UMAP, as densMAP more faithfully preserves the local density structure in the low-dimensional representation (see Figure \ref{fig:3:merging_merger_fraction_densmap}). 

Mutual information for discrete random variables, $X,\,Y,$ is defined by
\begin{equation}
    I(X;Y)=\sum_{x\in X}\sum_{y\in Y}P_{X,Y}(x,y)\log\bigg(\frac{P_{X,Y}(x,y)}{P_X(x)\cdot P_Y(y)}\bigg),
\end{equation}
where $P_{X,Y}$ is the joint probability function and $P_X,\,P_Y$ are the marginal probability functions, respectively \citep{shannon_mathematical_1948}. In our case, the histograms we have are given by the distribution $P(\mu\, | D_1, D_2)$, where $\mu$ can be any parameter associated with each image cutout and $D_1,\,D_2$ are the densMAP axes. Each bin of the histogram represents the mean of $\mu$ for the samples in the given bin. For each $\mu$ we obtain a $P_\mu(D_1,D_2)$ by shifting $\mu$ so that all values are positive, and we normalized the histogram to
\begin{equation}
    P_\mu(D_1,D_2)\equiv\frac{P(\mu\,|D_1,D_2)}{\sum_{d_1\in D_1}\sum_{d_2\in D_2}P(\mu\,|\,d_1,d_2)}.
\end{equation}
Note that this operation does not strictly yield a valid probability distribution; we used the normalized distribution solely to enable the computation of mutual information and to quantify the structure in the densMAPs. The mutual information is therefore
\begin{equation}
\begin{split}
    I_\mu(D_1;D_2) &= \sum_{d_1 \in D_1} \sum_{d_2 \in D_2} \bigg[ P_{\mu}(d_1,d_2) \times\,...\\
    ...&\times \log\bigg(\frac{P_\mu(d_1,d_2)}{P_{\mu,D_1}(d_1)\,P_{\mu,D_2}(d_2)}\bigg)\bigg],
\end{split}
\end{equation}
which we obtain for every $\mu$ that we have. 

We estimated the uncertainty in the mutual information using a Monte Carlo bootstrap procedure. For each parameter, $\mu$, we generated 5000 synthetic histograms. Each synthetic histogram was constructed by resampling every bin from a normal distribution whose mean and standard deviation are given by that bin’s mean and standard deviation. Bins with a single count are clipped. We computed the mutual information for each synthetic histogram using the method described above and took the standard deviation of the resulting distribution of mutual information values as the Monte Carlo bootstrap error estimate.

Figure \ref{fig:3:merging_merger_fraction_densmap} compares the densMAP projections of the latent representations from the models trained with tiered sigmoid and arcsinh scaling, each visualized as a mean histogram of the \texttt{merging\_merger\_fraction} parameter from the Galaxy Zoo DECaLS project. The tiered sigmoid scaling yields more significant mutual information, indicating a higher degree of clustering between the two densMAP axes and therefore more structure in the densMAP space when binned by \texttt{merging\_merger\_fraction}.

\begin{figure}
    \centering
    \includegraphics[width=\linewidth]{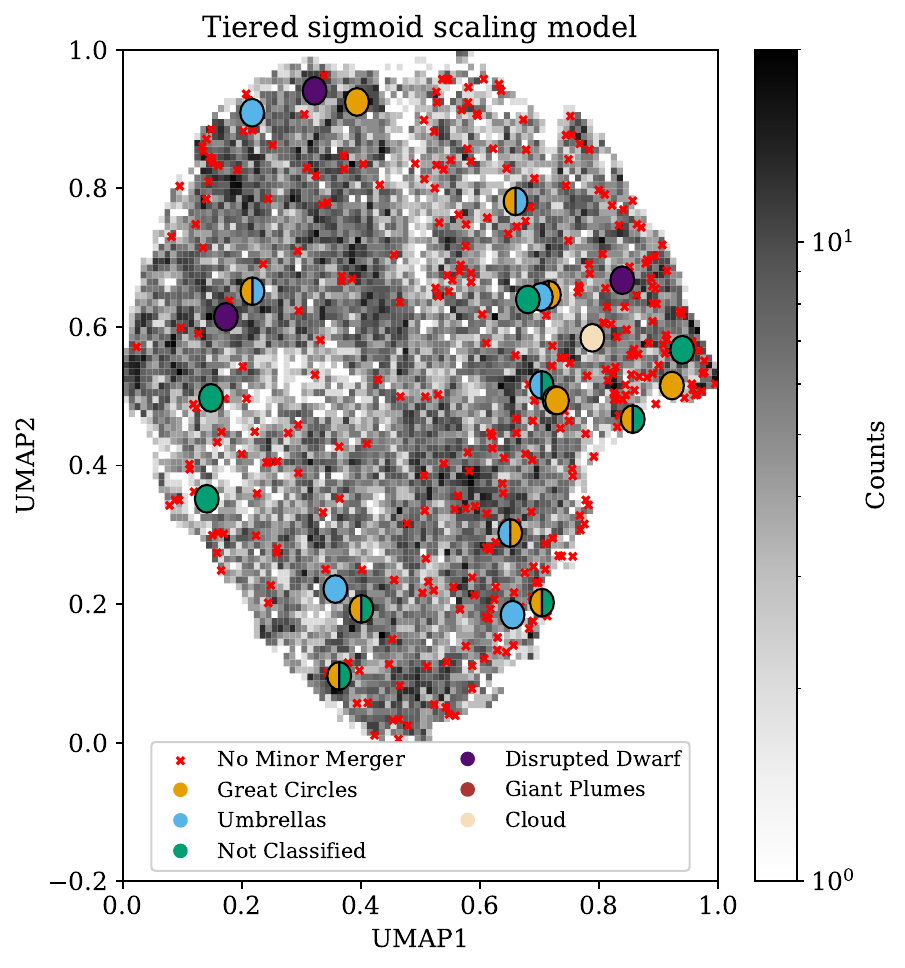}
    \caption{Labeled dataset of \cite{Miro-Carretero2024} passed through the tiered sigmoid scaling NNCLR encoder and projected to two dimensions by UMAP. The red crosses represent the negatives (inspected by \cite{Miro-Carretero2024} and found to have no low-surface-brightness tidal features), whereas the positives are marked by shaded circles using the morphologic classification described in \cite{Miro-Carretero2024}. The difficulty in classification becomes apparent with the high degree of scatter for both the positives and negatives.}
    \label{fig:3:labeled_set_umap}
\end{figure}

\begin{figure}
    \centering
    \begin{tabular}{c}
        \includegraphics[width=\columnwidth]{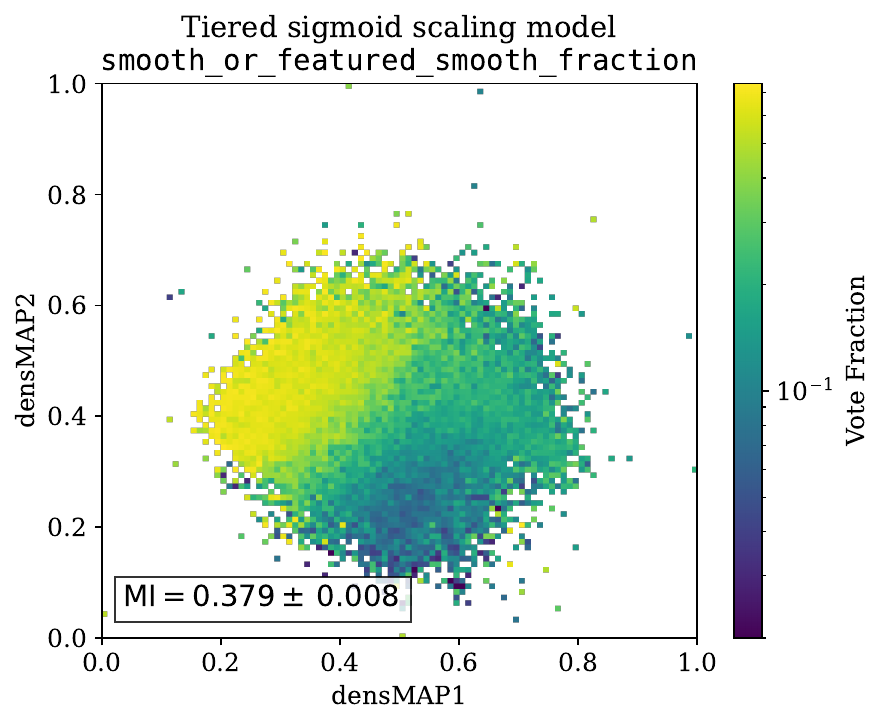} \\
        \includegraphics[width=\columnwidth]{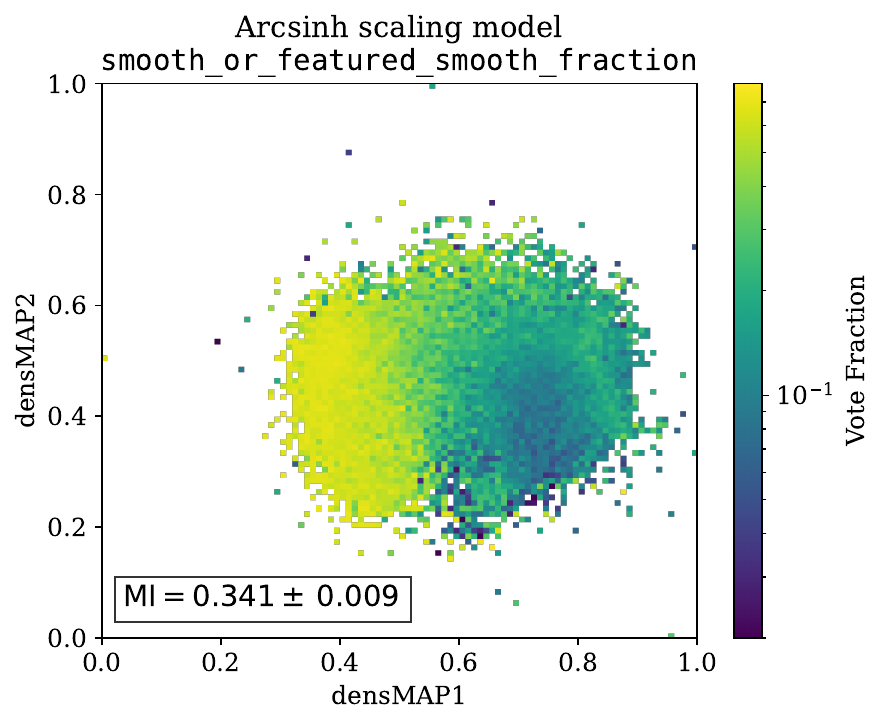}
    \end{tabular}
    \caption{(Top) Cross-matched Galaxy Zoo DECaLS \texttt{smooth\_or\_featured\_smooth\_fraction} reveals a morphological split determined by the NNCLR encoder. Notably, galactic outskirts retain sufficient information to reveal morphological categorizations. (Bottom) Latent representations of the model trained with arcsinh-scaled images replicates similar (although weaker, per the mutual information) clustering based on the aforementioned Galaxy Zoo DECaLS parameter.}
    \label{fig:3:smooth_or_featured_smooth_fraction_histograms}
\end{figure}

Having observed a region of the UMAP and densMAP relevant for the detection of tidal features, we next investigated whether the NNCLR encoder is able to retain information from low-surface-brightness features. Figure \ref{fig:3:labeled_set_umap} shows the embeddings obtained from the labeled set of stellar streams in \cite{Miro-Carretero2024}, along with their respective classifications (see Table A.1 in \cite{Miro-Carretero2024}). The wide dispersion of the labeled stream embeddings implies that low-surface-brightness features, while present, do not exert a sufficiently dominant influence on the organization of the NNCLR encoder's latent space as captured by the UMAP projection to represent the morphological scheme of \cite{Miro-Carretero2024}. Downstream classification tasks are therefore difficult to train successfully with the labeled dataset of this work without overfitting. We attempted to perform a binary classification task using the combined set of all images in Figure \ref{fig:3:labeled_set_umap} that are labeled as including a tidal feature; however, the results are not reliable.

Figure \ref{fig:3:smooth_or_featured_smooth_fraction_histograms} illustrates the correlations between the densMAP coordinates and the Galaxy Zoo DECaLS \texttt{smooth\_or\_featured\_smooth\_fraction}, which quantifies the fraction of citizen scientists identifying a galaxy as smooth rather than featured. We find a strong correspondence between higher smooth fraction values and one side of the densMAP projections in both models. This effect further reinforces the difference we observe between the second panel and the third and fourth panels of the UMAP mosaic seen in Figure \ref{fig:3:fiducial_mosaic}; that is, elliptical galaxies have a smoother transition in surface-brightness space to the background compared to other morphological types. The observed structure in both densMAP projections indicates that the NNCLR encoder effectively organizes galaxies according to visually salient properties of galactic outskirts for the  two models.

Although the tiered sigmoid scaling model arranges the latent space of representations according to various physical properties of the host galaxy and its environment, we found that this organization is not primarily driven by the tidal features of minor mergers as those identified by \cite{Miro-Carretero2024}. The tiered sigmoid scaling function suppresses dominant galactic features, but this suppression is incomplete since bright features continue to influence the latent space, as shown in Figure \ref{fig:3:smooth_or_featured_smooth_fraction_histograms} \citep[for a discussion of feature suppression in the context of contrastive learning, see][]{chen_intriguing_2021}. Achieving a latent space shaped purely by low-surface-brightness features would require stronger feature suppression and more targeted attention steering of the NNCLR encoder.

\subsection{Nearest neighbors of query images}
\begin{figure*}[h]
    \centering
    \begin{tabular}{c}
        \includegraphics[width=\textwidth]{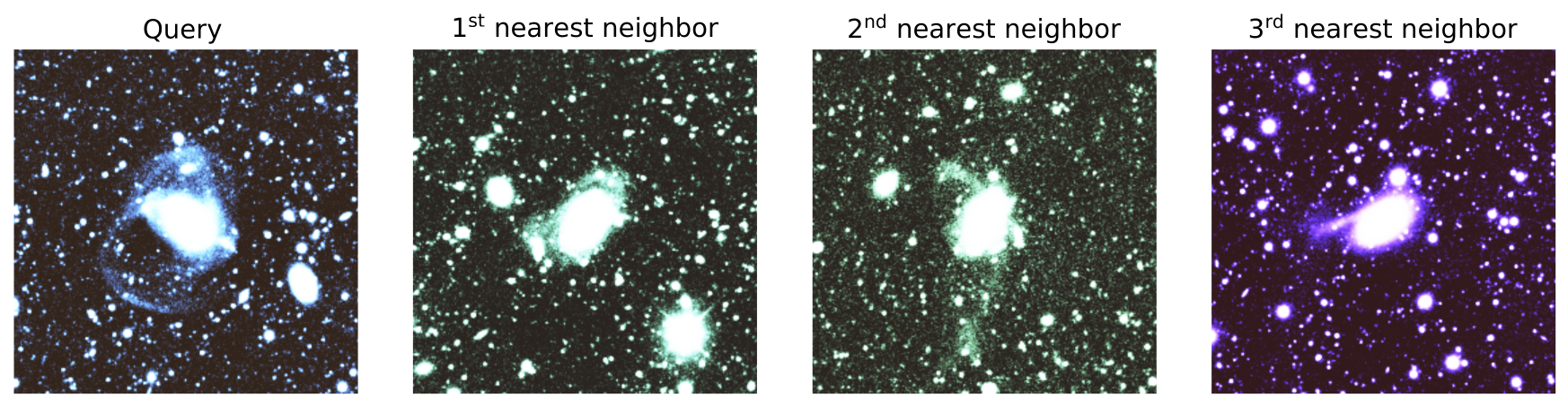} \\
        \includegraphics[width=\textwidth]{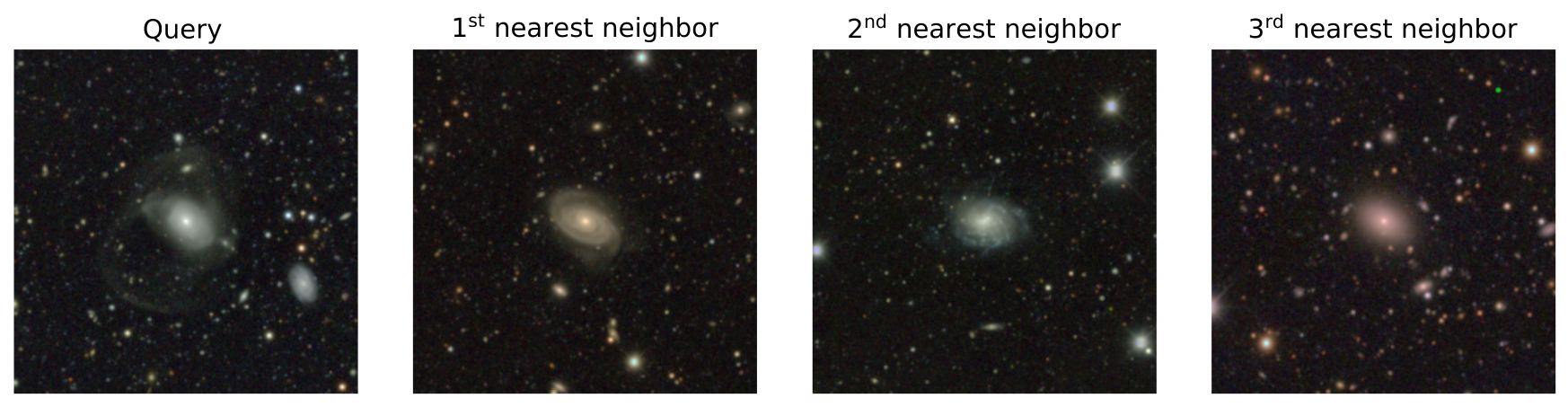}
    \end{tabular}
    \caption{Further probing of the latent space of representations of both our models by inputting a query image and checking for the $N$ nearest neighbors using the Euclidean distance. The query image in this case is LEDA 129430, which from visual inspection of our dataset appears to contain a candidate extragalactic stellar stream. The top row demonstrates the query image and its three nearest neighbors in the latent space of the model trained with the tiered-sigmoid-scaled images. The bottom row shows the nearest neighbors in the latent space of the model trained with the arcsinh-scaled images.}
    \label{fig:3:nearest_neighbors_of_sample}
\end{figure*}

We investigated the nearest neighbors in the latent space of representations of a galaxy containing low-surface-brightness features (LEDA 129430) in Figure \ref{fig:3:nearest_neighbors_of_sample}, where we compare those found using the tiered sigmoid scaling model (top) and the arcsinh scaling model (bottom). We define the nearest neighbors as those minimizing the Euclidean distance between the $\ell_2$ normalized latent representation of the query galaxy and those of other galaxies in the dataset.

Although low-surface-brightness features are present in the nearest-neighbor samples of both query images for each model, the low-surface-brightness features in the tiered-sigmoid-scaled images are easier to identify immediately and appear to have a tidal origin, in contrast to the neighbors obtained via the arcsinh scaling model. This highlights the benefit of a scaling function that is generic in emphasizing low-surface-brightness features without the need to manually tune the parameters of the scaling function on a per-sample basis.

\section{Conclusion}
In this work we implemented a self-supervised framework for characterizing low-surface-brightness tidal features in wide-field imaging by applying the NNCLR algorithm to DES DR2 data. Through the introduction of a tiered sigmoid scaling function, we effectively redirected the neural network’s implicit attention from the bright galactic centers toward their faint outskirts, where tidal substructures such as stellar streams are most likely to reside. Image preprocessing using the tiered sigmoid scaling function, combined with custom image-augmentation techniques for contrastive representation learning, produced embeddings with a higher degree of clustering of major-merger features (as classified by Galaxy Zoo DECaLS) in the DES DR2 sample.

Despite these advances, our analysis reveals that our implementation of self-supervised contrastive learning is insufficient to cleanly separate stellar streams in the latent space without further guidance. The dispersion of labeled examples from \cite{Miro-Carretero2024} indicates that additional supervision, more complex image scaling methods, or further machine-learning extensions will be necessary to enhance sensitivity to such faint, morphologically diverse features. Nonetheless, the resulting saliency maps confirm that the network learns to emphasize the flux variations most relevant to the identification of tidal debris, validating the conceptual design of the tiered sigmoid scaling.

Our results establish a methodological foundation for future applications of contrastive learning to deep survey data. The pipeline presented here may be extended to upcoming datasets from LSST and Euclid, providing an efficient, interpretable means of pretraining models on vast unlabeled image archives. In conjunction with small, curated labeled samples, this approach offers a scalable pathway toward the reliable automated detection of extragalactic stellar streams and other low-surface-brightness phenomena.

\begin{acknowledgements}
This work is financed by the Dutch Research Council (NWO) grant number OCENW.XL21.XL21.025. We thank the referee for their constructive comments. We thank the Center for Information Technology of the University of Groningen for their support and for providing access to the Hábrók high performance computing cluster. EBW thanks Ewoud Wempe for helpful discussions. AH thanks NWO for a Spinoza prize (SPI-78411). This research uses services or data provided by the Astro Data Lab, which is part of the Community Science and Data Center (CSDC) Program of NSF NOIRLab \citep{fitzpatrick_noao_2014, nikutta_data_2020}. NOIRLab is operated by the Association of Universities for Research in Astronomy (AURA), Inc. under a cooperative agreement with the U.S. National Science Foundation.
\end{acknowledgements}

\bibliographystyle{bibtex/aa}
\bibliography{bibtex/references}

@ARTICLE{Bertin1996,
       author = {{Bertin}, E. and {Arnouts}, S.},
        title = "{SExtractor: Software for source extraction.}",
      journal = {\aaps},
         year = "1996",
        month = "Jun",
       volume = {117},
        pages = {393-404},
          doi = {10.1051/aas:1996164},
       adsurl = {https://ui.adsabs.harvard.edu/abs/1996A&AS..117..393B}
}

@inproceedings{Adebayo2018,
 author = {Adebayo, Julius and Gilmer, Justin and Muelly, Michael and Goodfellow, Ian and Hardt, Moritz and Kim, Been},
 booktitle = {Advances in Neural Information Processing Systems},
 pages = {},
 title = {Sanity Checks for Saliency Maps},
 url = {https://proceedings.neurips.cc/paper_files/paper/2018/file/294a8ed24b1ad22ec2e7efea049b8737-Paper.pdf},
 volume = {31},
 year = {2018}
}

@article{Atkinson2013,
doi = {10.1088/0004-637X/765/1/28},
url = {https://dx.doi.org/10.1088/0004-637X/765/1/28},
year = {2013},
month = {feb},
publisher = {The American Astronomical Society},
volume = {765},
number = {1},
pages = {28},
author = {Atkinson, Adam M. and Abraham, Roberto G. and Ferguson, Annette M. N.},
title = {FAINT TIDAL FEATURES IN GALAXIES WITHIN THE CANADA–FRANCE–HAWAII TELESCOPE LEGACY SURVEY WIDE FIELDS},
journal = {\apj}
}

@article{Bovy2016,
doi = {10.3847/1538-4357/833/1/31},
url = {https://dx.doi.org/10.3847/1538-4357/833/1/31},
year = {2016},
month = {dec},
publisher = {The American Astronomical Society},
volume = {833},
number = {1},
pages = {31},
author = {Bovy, Jo and Bahmanyar, Anita and Fritz, Tobias K. and Kallivayalil, Nitya},
title = {THE SHAPE OF THE INNER MILKY WAY HALO FROM OBSERVATIONS OF THE PAL 5 AND GD–1 STELLAR STREAMS},
journal = {\apj}
}

@ARTICLE{Bowden2015,
       author = {{Bowden}, A. and {Belokurov}, V. and {Evans}, N.~W.},
        title = "{Dipping our toes in the water: first models of GD-1 as a stream}",
      journal = {\mnras},
         year = 2015,
        month = may,
       volume = {449},
       number = {2},
        pages = {1391-1400},
          doi = {10.1093/mnras/stv285},
archivePrefix = {arXiv},
       eprint = {1502.00484},
 primaryClass = {astro-ph.GA},
       adsurl = {https://ui.adsabs.harvard.edu/abs/2015MNRAS.449.1391B}
}

@ARTICLE{DES_DR1,
       author = {{Abbott}, T.~M.~C. and {Abdalla}, F.~B. and {Allam}, S. and {Amara}, A. and {Annis}, J. and {Asorey}, J. and {Avila}, S. and {Ballester}, O. and {Banerji}, M. and {Barkhouse}, W. and {Baruah}, L. and {Baumer}, M. and {Bechtol}, K. and {Becker}, M.~R. and {Benoit-L{\'e}vy}, A. and {Bernstein}, G.~M. and {Bertin}, E. and {Blazek}, J. and {Bocquet}, S. and {Brooks}, D. and {Brout}, D. and {Buckley-Geer}, E. and {Burke}, D.~L. and {Busti}, V. and {Campisano}, R. and {Cardiel-Sas}, L. and {Carnero Rosell}, A. and {Carrasco Kind}, M. and {Carretero}, J. and {Castander}, F.~J. and {Cawthon}, R. and {Chang}, C. and {Chen}, X. and {Conselice}, C. and {Costa}, G. and {Crocce}, M. and {Cunha}, C.~E. and {D'Andrea}, C.~B. and {da Costa}, L.~N. and {Das}, R. and {Daues}, G. and {Davis}, T.~M. and {Davis}, C. and {De Vicente}, J. and {DePoy}, D.~L. and {DeRose}, J. and {Desai}, S. and {Diehl}, H.~T. and {Dietrich}, J.~P. and {Dodelson}, S. and {Doel}, P. and {Drlica-Wagner}, A. and {Eifler}, T.~F. and {Elliott}, A.~E. and {Evrard}, A.~E. and {Farahi}, A. and {Fausti Neto}, A. and {Fernandez}, E. and {Finley}, D.~A. and {Flaugher}, B. and {Foley}, R.~J. and {Fosalba}, P. and {Friedel}, D.~N. and {Frieman}, J. and {Garc{\'\i}a-Bellido}, J. and {Gaztanaga}, E. and {Gerdes}, D.~W. and {Giannantonio}, T. and {Gill}, M.~S.~S. and {Glazebrook}, K. and {Goldstein}, D.~A. and {Gower}, M. and {Gruen}, D. and {Gruendl}, R.~A. and {Gschwend}, J. and {Gupta}, R.~R. and {Gutierrez}, G. and {Hamilton}, S. and {Hartley}, W.~G. and {Hinton}, S.~R. and {Hislop}, J.~M. and {Hollowood}, D. and {Honscheid}, K. and {Hoyle}, B. and {Huterer}, D. and {Jain}, B. and {James}, D.~J. and {Jeltema}, T. and {Johnson}, M.~W.~G. and {Johnson}, M.~D. and {Kacprzak}, T. and {Kent}, S. and {Khullar}, G. and {Klein}, M. and {Kovacs}, A. and {Koziol}, A.~M.~G. and {Krause}, E. and {Kremin}, A. and {Kron}, R. and {Kuehn}, K. and {Kuhlmann}, S. and {Kuropatkin}, N. and {Lahav}, O. and {Lasker}, J. and {Li}, T.~S. and {Li}, R.~T. and {Liddle}, A.~R. and {Lima}, M. and {Lin}, H. and {L{\'o}pez-Reyes}, P. and {MacCrann}, N. and {Maia}, M.~A.~G. and {Maloney}, J.~D. and {Manera}, M. and {March}, M. and {Marriner}, J. and {Marshall}, J.~L. and {Martini}, P. and {McClintock}, T. and {McKay}, T. and {McMahon}, R.~G. and {Melchior}, P. and {Menanteau}, F. and {Miller}, C.~J. and {Miquel}, R. and {Mohr}, J.~J. and {Morganson}, E. and {Mould}, J. and {Neilsen}, E. and {Nichol}, R.~C. and {Nogueira}, F. and {Nord}, B. and {Nugent}, P. and {Nunes}, L. and {Ogando}, R.~L.~C. and {Old}, L. and {Pace}, A.~B. and {Palmese}, A. and {Paz-Chinch{\'o}n}, F. and {Peiris}, H.~V. and {Percival}, W.~J. and {Petravick}, D. and {Plazas}, A.~A. and {Poh}, J. and {Pond}, C. and {Porredon}, A. and {Pujol}, A. and {Refregier}, A. and {Reil}, K. and {Ricker}, P.~M. and {Rollins}, R.~P. and {Romer}, A.~K. and {Roodman}, A. and {Rooney}, P. and {Ross}, A.~J. and {Rykoff}, E.~S. and {Sako}, M. and {Sanchez}, M.~L. and {Sanchez}, E. and {Santiago}, B. and {Saro}, A. and {Scarpine}, V. and {Scolnic}, D. and {Serrano}, S. and {Sevilla-Noarbe}, I. and {Sheldon}, E. and {Shipp}, N. and {Silveira}, M.~L. and {Smith}, M. and {Smith}, R.~C. and {Smith}, J.~A. and {Soares-Santos}, M. and {Sobreira}, F. and {Song}, J. and {Stebbins}, A. and {Suchyta}, E. and {Sullivan}, M. and {Swanson}, M.~E.~C. and {Tarle}, G. and {Thaler}, J. and {Thomas}, D. and {Thomas}, R.~C. and {Troxel}, M.~A. and {Tucker}, D.~L. and {Vikram}, V. and {Vivas}, A.~K. and {Walker}, A.~R. and {Wechsler}, R.~H. and {Weller}, J. and {Wester}, W. and {Wolf}, R.~C. and {Wu}, H. and {Yanny}, B. and {Zenteno}, A. and {Zhang}, Y. and {Zuntz}, J. and {DES Collaboration} and {Juneau}, S. and {Fitzpatrick}, M. and {Nikutta}, R.},
        title = "{The Dark Energy Survey: Data Release 1}",
      journal = {\apjs},
         year = 2018,
        month = dec,
       volume = {239},
       number = {2},
          eid = {18},
        pages = {18},
          doi = {10.3847/1538-4365/aae9f0},
archivePrefix = {arXiv},
       eprint = {1801.03181},
 primaryClass = {astro-ph.IM},
       adsurl = {https://ui.adsabs.harvard.edu/abs/2018ApJS..239...18A}
}

@article{DES_DR2,
   title={The Dark Energy Survey Data Release 2},
   volume={255},
   ISSN={1538-4365},
   url={http://dx.doi.org/10.3847/1538-4365/ac00b3},
   DOI={10.3847/1538-4365/ac00b3},
   number={2},
   journal={\apjs},
   publisher={American Astronomical Society},
   author={Abbott, T. M. C. and Adamów, M. and Aguena, M. and Allam, S. and Amon, A. and Annis, J. and Avila, S. and Bacon, D. and Banerji, M. and Bechtol, K. and Becker, M. R. and Bernstein, G. M. and Bertin, E. and Bhargava, S. and Bridle, S. L. and Brooks, D. and Burke, D. L. and Carnero Rosell, A. and Carrasco Kind, M. and Carretero, J. and Castander, F. J. and Cawthon, R. and Chang, C. and Choi, A. and Conselice, C. and Costanzi, M. and Crocce, M. and da Costa, L. N. and Davis, T. M. and De Vicente, J. and DeRose, J. and Desai, S. and Diehl, H. T. and Dietrich, J. P. and Drlica-Wagner, A. and Eckert, K. and Elvin-Poole, J. and Everett, S. and Evrard, A. E. and Ferrero, I. and Ferté, A. and Flaugher, B. and Fosalba, P. and Friedel, D. and Frieman, J. and García-Bellido, J. and Gaztanaga, E. and Gelman, L. and Gerdes, D. W. and Giannantonio, T. and Gill, M. S. S. and Gruen, D. and Gruendl, R. A. and Gschwend, J. and Gutierrez, G. and Hartley, W. G. and Hinton, S. R. and Hollowood, D. L. and Honscheid, K. and Huterer, D. and James, D. J. and Jeltema, T. and Johnson, M. D. and Kent, S. and Kron, R. and Kuehn, K. and Kuropatkin, N. and Lahav, O. and Li, T. S. and Lidman, C. and Lin, H. and MacCrann, N. and Maia, M. A. G. and Manning, T. A. and Maloney, J. D. and March, M. and Marshall, J. L. and Martini, P. and Melchior, P. and Menanteau, F. and Miquel, R. and Morgan, R. and Myles, J. and Neilsen, E. and Ogando, R. L. C. and Palmese, A. and Paz-Chinchón, F. and Petravick, D. and Pieres, A. and Plazas, A. A. and Pond, C. and Rodriguez-Monroy, M. and Romer, A. K. and Roodman, A. and Rykoff, E. S. and Sako, M. and Sanchez, E. and Santiago, B. and Scarpine, V. and Serrano, S. and Sevilla-Noarbe, I. and Smith, J. Allyn and Smith, M. and Soares-Santos, M. and Suchyta, E. and Swanson, M. E. C. and Tarle, G. and Thomas, D. and To, C. and Tremblay, P. E. and Troxel, M. A. and Tucker, D. L. and Turner, D. J. and Varga, T. N. and Walker, A. R. and Wechsler, R. H. and Weller, J. and Wester, W. and Wilkinson, R. D. and Yanny, B. and Zhang, Y. and Nikutta, R. and Fitzpatrick, M. and Jacques, A. and Scott, A. and Olsen, K. and Huang, L. and Herrera, D. and Juneau, S. and Nidever, D. and Weaver, B. A. and Adean, C. and Correia, V. and de Freitas, M. and Freitas, F. N. and Singulani, C. and Vila-Verde, G.},
   year={2021},
   month=jul, pages={20} }

@ARTICLE{Desmons2024,
       author = {{Desmons}, Alice and {Brough}, Sarah and {Lanusse}, Francois},
        title = "{Detecting galaxy tidal features using self-supervised representation learning}",
      journal = {\mnras},
         year = 2024,
        month = jul,
       volume = {531},
       number = {4},
        pages = {4070-4084},
          doi = {10.1093/mnras/stae1402},
archivePrefix = {arXiv},
       eprint = {2308.07962},
 primaryClass = {astro-ph.GA},
       adsurl = {https://ui.adsabs.harvard.edu/abs/2024MNRAS.531.4070D}
}

@article{Dey2019,
doi = {10.3847/1538-3881/ab089d},
url = {https://dx.doi.org/10.3847/1538-3881/ab089d},
year = {2019},
month = {apr},
publisher = {The American Astronomical Society},
volume = {157},
number = {5},
pages = {168},
author = {Dey, Arjun and Schlegel, David J. and Lang, Dustin and Blum, Robert and Burleigh, Kaylan and Fan, Xiaohui and Findlay, Joseph R. and Finkbeiner, Doug and Herrera, David and Juneau, Stéphanie and Landriau, Martin and Levi, Michael and McGreer, Ian and Meisner, Aaron and Myers, Adam D. and Moustakas, John and Nugent, Peter and Patej, Anna and Schlafly, Edward F. and Walker, Alistair R. and Valdes, Francisco and Weaver, Benjamin A. and Yèche, Christophe and Zou, Hu and Zhou, Xu and Abareshi, Behzad and Abbott, T. M. C. and Abolfathi, Bela and Aguilera, C. and Alam, Shadab and Allen, Lori and Alvarez, A. and Annis, James and Ansarinejad, Behzad and Aubert, Marie and Beechert, Jacqueline and Bell, Eric F. and BenZvi, Segev Y. and Beutler, Florian and Bielby, Richard M. and Bolton, Adam S. and Briceño, César and Buckley-Geer, Elizabeth J. and Butler, Karen and Calamida, Annalisa and Carlberg, Raymond G. and Carter, Paul and Casas, Ricard and Castander, Francisco J. and Choi, Yumi and Comparat, Johan and Cukanovaite, Elena and Delubac, Timothée and DeVries, Kaitlin and Dey, Sharmila and Dhungana, Govinda and Dickinson, Mark and Ding, Zhejie and Donaldson, John B. and Duan, Yutong and Duckworth, Christopher J. and Eftekharzadeh, Sarah and Eisenstein, Daniel J. and Etourneau, Thomas and Fagrelius, Parker A. and Farihi, Jay and Fitzpatrick, Mike and Font-Ribera, Andreu and Fulmer, Leah and Gänsicke, Boris T. and Gaztanaga, Enrique and George, Koshy and Gerdes, David W. and A Gontcho, Satya Gontcho and Gorgoni, Claudio and Green, Gregory and Guy, Julien and Harmer, Diane and Hernandez, M. and Honscheid, Klaus and Huang, Lijuan (Wendy) and James, David J. and Jannuzi, Buell T. and Jiang, Linhua and Joyce, Richard and Karcher, Armin and Karkar, Sonia and Kehoe, Robert and Kneib, Jean-Paul and Kueter-Young, Andrea and Lan, Ting-Wen and Lauer, Tod R. and Guillou, Laurent Le and Van Suu, Auguste Le and Lee, Jae Hyeon and Lesser, Michael and Levasseur, Laurence Perreault and Li, Ting S. and Mann, Justin L. and Marshall, Robert and Martínez-Vázquez, C. E. and Martini, Paul and du Mas des Bourboux, Hélion and McManus, Sean and Meier, Tobias Gabriel and Ménard, Brice and Metcalfe, Nigel and Muñoz-Gutiérrez, Andrea and Najita, Joan and Napier, Kevin and Narayan, Gautham and Newman, Jeffrey A. and Nie, Jundan and Nord, Brian and Norman, Dara J. and Olsen, Knut A. G. and Paat, Anthony and Palanque-Delabrouille, Nathalie and Peng, Xiyan and Poppett, Claire L. and Poremba, Megan R. and Prakash, Abhishek and Rabinowitz, David and Raichoor, Anand and Rezaie, Mehdi and Robertson, A. N. and Roe, Natalie A. and Ross, Ashley J. and Ross, Nicholas P. and Rudnick, Gregory and Gaines, Sasha and Saha, Abhijit and Sánchez, F. Javier and Savary, Elodie and Schweiker, Heidi and Scott, Adam and Seo, Hee-Jong and Shan, Huanyuan and Silva, David R. and Slepian, Zachary and Soto, Christian and Sprayberry, David and Staten, Ryan and Stillman, Coley M. and Stupak, Robert J. and Summers, David L. and Tie, Suk Sien and Tirado, H. and Vargas-Magaña, Mariana and Vivas, A. Katherina and Wechsler, Risa H. and Williams, Doug and Yang, Jinyi and Yang, Qian and Yapici, Tolga and Zaritsky, Dennis and Zenteno, A. and Zhang, Kai and Zhang, Tianmeng and Zhou, Rongpu and Zhou, Zhimin},
title = {Overview of the DESI Legacy Imaging Surveys},
journal = {\aj}
}

@inproceedings{Dosovitskiy2021,
    title={An Image is Worth 16x16 Words: Transformers for Image Recognition at Scale},
    author={Alexey Dosovitskiy and Lucas Beyer and Alexander Kolesnikov and Dirk Weissenborn and Xiaohua Zhai and Thomas Unterthiner and Mostafa Dehghani and Matthias Minderer and Georg Heigold and Sylvain Gelly and Jakob Uszkoreit and Neil Houlsby},
    booktitle={International Conference on Learning Representations},
    year={2021},
    url={https://openreview.net/forum?id=YicbFdNTTy}
}

@INPROCEEDINGS{Dwibedi2021,
author = { Dwibedi, Debidatta and Aytar, Yusuf and Tompson, Jonathan and Sermanet, Pierre and Zisserman, Andrew },
booktitle = { 2021 IEEE/CVF International Conference on Computer Vision (ICCV) },
title = {{ With a Little Help from My Friends: Nearest-Neighbor Contrastive Learning of Visual Representations }},
year = {2021},
volume = {},
ISSN = {},
pages = {9568-9577},
doi = {10.1109/ICCV48922.2021.00945},
url = {https://doi.ieeecomputersociety.org/10.1109/ICCV48922.2021.00945},
publisher = {IEEE Computer Society},
month =Oct}

@ARTICLE{EuclidWideSurvey,
       author = {{Euclid Collaboration} and {Scaramella}, R. and {Amiaux}, J. and {Mellier}, Y. and {Burigana}, C. and {Carvalho}, C.~S. and {Cuillandre}, J. -C. and {Da Silva}, A. and {Derosa}, A. and {Dinis}, J. and {Maiorano}, E. and {Maris}, M. and {Tereno}, I. and {Laureijs}, R. and {Boenke}, T. and {Buenadicha}, G. and {Dupac}, X. and {Gaspar Venancio}, L.~M. and {G{\'o}mez-{\'A}lvarez}, P. and {Hoar}, J. and {Lorenzo Alvarez}, J. and {Racca}, G.~D. and {Saavedra-Criado}, G. and {Schwartz}, J. and {Vavrek}, R. and {Schirmer}, M. and {Aussel}, H. and {Azzollini}, R. and {Cardone}, V.~F. and {Cropper}, M. and {Ealet}, A. and {Garilli}, B. and {Gillard}, W. and {Granett}, B.~R. and {Guzzo}, L. and {Hoekstra}, H. and {Jahnke}, K. and {Kitching}, T. and {Maciaszek}, T. and {Meneghetti}, M. and {Miller}, L. and {Nakajima}, R. and {Niemi}, S.~M. and {Pasian}, F. and {Percival}, W.~J. and {Pottinger}, S. and {Sauvage}, M. and {Scodeggio}, M. and {Wachter}, S. and {Zacchei}, A. and {Aghanim}, N. and {Amara}, A. and {Auphan}, T. and {Auricchio}, N. and {Awan}, S. and {Balestra}, A. and {Bender}, R. and {Bodendorf}, C. and {Bonino}, D. and {Branchini}, E. and {Brau-Nogue}, S. and {Brescia}, M. and {Candini}, G.~P. and {Capobianco}, V. and {Carbone}, C. and {Carlberg}, R.~G. and {Carretero}, J. and {Casas}, R. and {Castander}, F.~J. and {Castellano}, M. and {Cavuoti}, S. and {Cimatti}, A. and {Cledassou}, R. and {Congedo}, G. and {Conselice}, C.~J. and {Conversi}, L. and {Copin}, Y. and {Corcione}, L. and {Costille}, A. and {Courbin}, F. and {Degaudenzi}, H. and {Douspis}, M. and {Dubath}, F. and {Duncan}, C.~A.~J. and {Dusini}, S. and {Farrens}, S. and {Ferriol}, S. and {Fosalba}, P. and {Fourmanoit}, N. and {Frailis}, M. and {Franceschi}, E. and {Franzetti}, P. and {Fumana}, M. and {Gillis}, B. and {Giocoli}, C. and {Grazian}, A. and {Grupp}, F. and {Haugan}, S.~V.~H. and {Holmes}, W. and {Hormuth}, F. and {Hudelot}, P. and {Kermiche}, S. and {Kiessling}, A. and {Kilbinger}, M. and {Kohley}, R. and {Kubik}, B. and {K{\"u}mmel}, M. and {Kunz}, M. and {Kurki-Suonio}, H. and {Lahav}, O. and {Ligori}, S. and {Lilje}, P.~B. and {Lloro}, I. and {Mansutti}, O. and {Marggraf}, O. and {Markovic}, K. and {Marulli}, F. and {Massey}, R. and {Maurogordato}, S. and {Melchior}, M. and {Merlin}, E. and {Meylan}, G. and {Mohr}, J.~J. and {Moresco}, M. and {Morin}, B. and {Moscardini}, L. and {Munari}, E. and {Nichol}, R.~C. and {Padilla}, C. and {Paltani}, S. and {Peacock}, J. and {Pedersen}, K. and {Pettorino}, V. and {Pires}, S. and {Poncet}, M. and {Popa}, L. and {Pozzetti}, L. and {Raison}, F. and {Rebolo}, R. and {Rhodes}, J. and {Rix}, H. -W. and {Roncarelli}, M. and {Rossetti}, E. and {Saglia}, R. and {Schneider}, P. and {Schrabback}, T. and {Secroun}, A. and {Seidel}, G. and {Serrano}, S. and {Sirignano}, C. and {Sirri}, G. and {Skottfelt}, J. and {Stanco}, L. and {Starck}, J.~L. and {Tallada-Cresp{\'\i}}, P. and {Tavagnacco}, D. and {Taylor}, A.~N. and {Teplitz}, H.~I. and {Toledo-Moreo}, R. and {Torradeflot}, F. and {Trifoglio}, M. and {Valentijn}, E.~A. and {Valenziano}, L. and {Verdoes Kleijn}, G.~A. and {Wang}, Y. and {Welikala}, N. and {Weller}, J. and {Wetzstein}, M. and {Zamorani}, G. and {Zoubian}, J. and {Andreon}, S. and {Baldi}, M. and {Bardelli}, S. and {Boucaud}, A. and {Camera}, S. and {Di Ferdinando}, D. and {Fabbian}, G. and {Farinelli}, R. and {Galeotta}, S. and {Graci{\'a}-Carpio}, J. and {Maino}, D. and {Medinaceli}, E. and {Mei}, S. and {Neissner}, C. and {Polenta}, G. and {Renzi}, A. and {Romelli}, E. and {Rosset}, C. and {Sureau}, F. and {Tenti}, M. and {Vassallo}, T. and {Zucca}, E. and {Baccigalupi}, C. and {Balaguera-Antol{\'\i}nez}, A. and {Battaglia}, P. and {Biviano}, A. and {Borgani}, S. and {Bozzo}, E. and {Cabanac}, R. and {Cappi}, A.},
        title = "{Euclid preparation. I. The Euclid Wide Survey}",
      journal = {\aap},
         year = 2022,
        month = jun,
       volume = {662},
          eid = {A112},
        pages = {A112},
          doi = {10.1051/0004-6361/202141938},
archivePrefix = {arXiv},
       eprint = {2108.01201},
 primaryClass = {astro-ph.CO},
       adsurl = {https://ui.adsabs.harvard.edu/abs/2022A&A...662A.112E}
}

@ARTICLE{Gordon2024,
       author = {{Gordon}, Alexander J. and {Ferguson}, Annette M.~N. and {Mann}, Robert G.},
        title = "{Uncovering tidal treasures: automated classification of faint tidal features in DECaLS data}",
      journal = {\mnras},
         year = 2024,
        month = oct,
       volume = {534},
       number = {2},
        pages = {1459-1480},
          doi = {10.1093/mnras/stae2169},
archivePrefix = {arXiv},
       eprint = {2404.06487},
 primaryClass = {astro-ph.GA},
       adsurl = {https://ui.adsabs.harvard.edu/abs/2024MNRAS.534.1459G}
}

@ARTICLE{Helmi1999,
       author = {{Helmi}, Amina and {White}, Simon D.~M.},
        title = "{Building up the stellar halo of the Galaxy}",
      journal = {\mnras},
         year = 1999,
        month = aug,
       volume = {307},
       number = {3},
        pages = {495-517},
          doi = {10.1046/j.1365-8711.1999.02616.x},
archivePrefix = {arXiv},
       eprint = {astro-ph/9901102},
 primaryClass = {astro-ph},
       adsurl = {https://ui.adsabs.harvard.edu/abs/1999MNRAS.307..495H}
}

@INPROCEEDINGS{Kaiming2016,
       author = {{He}, Kaiming and {Zhang}, Xiangyu and {Ren}, Shaoqing and {Sun}, Jian},
        title = "{Deep Residual Learning for Image Recognition}",
    booktitle = {2016 IEEE Conference on Computer Vision and Pattern Recognition (CVPR)},
         year = 2016,
        month = jun,
          doi = {10.1109/CVPR.2016.90},
archivePrefix = {arXiv},
       eprint = {1512.03385},
 primaryClass = {cs.CV},
       adsurl = {https://ui.adsabs.harvard.edu/abs/2016cvpr.confE...1H}
}

@ARTICLE{Kobourov2012,
       author = {{Kobourov}, Stephen G.},
        title = "{Spring Embedders and Force Directed Graph Drawing Algorithms}",
      journal = {arXiv e-prints},
         year = 2012,
        month = jan,
          eid = {arXiv:1201.3011},
        pages = {arXiv:1201.3011},
          doi = {10.48550/arXiv.1201.3011},
archivePrefix = {arXiv},
       eprint = {1201.3011},
 primaryClass = {cs.CG},
       adsurl = {https://ui.adsabs.harvard.edu/abs/2012arXiv1201.3011K}
}

@ARTICLE{Koposov2010,
       author = {{Koposov}, Sergey E. and {Rix}, Hans-Walter and {Hogg}, David W.},
        title = "{Constraining the Milky Way Potential with a Six-Dimensional Phase-Space Map of the GD-1 Stellar Stream}",
      journal = {\apj},
         year = 2010,
        month = mar,
       volume = {712},
       number = {1},
        pages = {260-273},
          doi = {10.1088/0004-637X/712/1/260},
archivePrefix = {arXiv},
       eprint = {0907.1085},
 primaryClass = {astro-ph.GA},
       adsurl = {https://ui.adsabs.harvard.edu/abs/2010ApJ...712..260K}
}

@ARTICLE{Lintott2008,
       author = {{Lintott}, Chris J. and {Schawinski}, Kevin and {Slosar}, An{\v{z}}e and {Land}, Kate and {Bamford}, Steven and {Thomas}, Daniel and {Raddick}, M. Jordan and {Nichol}, Robert C. and {Szalay}, Alex and {Andreescu}, Dan and {Murray}, Phil and {Vandenberg}, Jan},
        title = "{Galaxy Zoo: morphologies derived from visual inspection of galaxies from the Sloan Digital Sky Survey}",
      journal = {\mnras},
         year = 2008,
        month = sep,
       volume = {389},
       number = {3},
        pages = {1179-1189},
          doi = {10.1111/j.1365-2966.2008.13689.x},
archivePrefix = {arXiv},
       eprint = {0804.4483},
 primaryClass = {astro-ph},
       adsurl = {https://ui.adsabs.harvard.edu/abs/2008MNRAS.389.1179L}
}

@inproceedings{Loshchilov2017,
    title={Decoupled Weight Decay Regularization},
    author={Ilya Loshchilov and Frank Hutter},
    booktitle={International Conference on Learning Representations},
    year={2019},
    url={https://openreview.net/forum?id=Bkg6RiCqY7},
}

@ARTICLE{Majewski1996,
       author = {{Majewski}, Steven R. and {Munn}, Jeffrey A. and {Hawley}, Suzanne L.},
        title = "{Absolute Proper Motions to B approximately 22.5: Large-Scale Streaming Motions and the Structure and Origin of the Galactic Halo}",
      journal = {\apjl},
         year = 1996,
        month = mar,
       volume = {459},
        pages = {L73},
          doi = {10.1086/309944},
       adsurl = {https://ui.adsabs.harvard.edu/abs/1996ApJ...459L..73M}
}

@article{Martinez-Delgado2025,
	author = {{Mart\'{\i}nez-Delgado}, David and {Stein}, Michael and {Sakowska}, Joanna D. and {Maurice Weigelt, M.} and {Rom\'an, Javier} and {Donatiello, Giuseppe} and {Roca-F\`abrega, Santi} and {Schirmer, Mischa} and {Grebel, Eva K.} and {Saifollahi, Teymoor} and {Kanipe, Jeff} and {G\'omez-Flechoso, M. Angeles} and {Akhlaghi, Mohammad} and {Javanmardi, Behnam} and {Wu, Gang} and {Eskandarlou, Sepideh} and {Bomans, Dominik J.} and {Henkel, Cristian} and {Block, Adam} and {Hanson, Mark} and {Schedler, Johannes} and {Teuwen, Karel} and {GaBany, R. Jay} and {Iba\~nez Perez, Alvaro} and {Crawford, Ken} and {Promper, Wolfgang} and {Jimenez, Manuel} and {Farr\`as-Aloy, S\'{\i}lvia} and {Mir\'o-Carretero, Juan}},
	title = {Stellar tidal streams around nearby spiral galaxies with deep imaging from amateur telescopes},
	DOI= "10.1051/0004-6361/202554980",
	url= "https://doi.org/10.1051/0004-6361/202554980",
	journal = {A\&A},
	year = 2025,
	volume = 701,
	pages = "A182",
}

@article{McInnes2018,
      title={UMAP: Uniform Manifold Approximation and Projection for Dimension Reduction}, 
      author={Leland McInnes and John Healy and James Melville},
      year={2018},
      eprint={1802.03426},
      pages={arXiv:1802.03426},
      journal = {arXiv e-prints},
      archivePrefix={arXiv},
      primaryClass={stat.ML},
      url={https://arxiv.org/abs/1802.03426}, 
}

@ARTICLE{Miro-Carretero2024,
       author = {{Mir{\'o}-Carretero}, Juan and {Mart{\'\i}nez-Delgado}, David and {G{\'o}mez-Flechoso}, Mar{\'\i}a A. and {Cooper}, Andrew and {Akhlaghi}, Mohammad and {Donatiello}, Giuseppe and {Kuijken}, Konrad and {Lang}, Dustin and {Makarov}, Dmitry and {Laine}, Seppo and {Roca-F{\`a}brega}, Santi},
        title = "{Extragalactic stellar tidal streams in the Dark Energy Survey}",
      journal = {\aap},
         year = 2024,
        month = nov,
       volume = {691},
          eid = {A196},
        pages = {A196},
          doi = {10.1051/0004-6361/202451685},
archivePrefix = {arXiv},
       eprint = {2407.20483},
 primaryClass = {astro-ph.GA},
       adsurl = {https://ui.adsabs.harvard.edu/abs/2024A&A...691A.196M}
}

@article{Sanchez2023,
    author = {Sánchez, H Domínguez and Martin, G and Damjanov, I and Buitrago, F and Huertas-Company, M and Bottrell, C and Bernardi, M and Knapen, J H and Vega-Ferrero, J and Hausen, R and Kado-Fong, E and Población-Criado, D and Souchereau, H and Leste, O K and Robertson, B and Sahelices, B and Johnston, K V},
    title = {Identification of tidal features in deep optical galaxy images with convolutional neural networks},
    journal = {\mnras},
    volume = {521},
    number = {3},
    pages = {3861-3872},
    year = {2023},
    month = {03},
    issn = {0035-8711},
    doi = {10.1093/mnras/stad750},
    url = {https://doi.org/10.1093/mnras/stad750},
    eprint = {https://academic.oup.com/mnras/article-pdf/521/3/3861/49632103/stad750.pdf},
}

@ARTICLE{Sola2022,
       author = {{Sola}, Elisabeth and {Duc}, Pierre-Alain and {Richards}, Felix and {Paiement}, Adeline and {Urbano}, Mathias and {Klehammer}, Julie and {B{\'\i}lek}, Michal and {Cuillandre}, Jean-Charles and {Gwyn}, Stephen and {McConnachie}, Alan},
        title = "{Characterization of low surface brightness structures in annotated deep images}",
      journal = {\aap},
         year = 2022,
        month = jun,
       volume = {662},
          eid = {A124},
        pages = {A124},
          doi = {10.1051/0004-6361/202142675},
archivePrefix = {arXiv},
       eprint = {2203.03973},
 primaryClass = {astro-ph.GA},
       adsurl = {https://ui.adsabs.harvard.edu/abs/2022A&A...662A.124S}
}

@ARTICLE{Schlegel1998,
       author = {{Schlegel}, David J. and {Finkbeiner}, Douglas P. and {Davis}, Marc},
        title = "{Maps of Dust Infrared Emission for Use in Estimation of Reddening and Cosmic Microwave Background Radiation Foregrounds}",
      journal = {\apj},
         year = 1998,
        month = jun,
       volume = {500},
       number = {2},
        pages = {525-553},
          doi = {10.1086/305772},
archivePrefix = {arXiv},
       eprint = {astro-ph/9710327},
 primaryClass = {astro-ph},
       adsurl = {https://ui.adsabs.harvard.edu/abs/1998ApJ...500..525S}
}

@article{Simonyan2014,
      title={Deep Inside Convolutional Networks: Visualising Image Classification Models and Saliency Maps}, 
      author={Karen Simonyan and Andrea Vedaldi and Andrew Zisserman},
      year={2014},
      journal={arXiv e-prints},
      eprint={1312.6034},
      pages={arXiv:1312.6034},
      archivePrefix={arXiv},
      primaryClass={cs.CV},
      url={https://arxiv.org/abs/1312.6034}, 
}

@ARTICLE{Stetson1987,
       author = {{Stetson}, Peter B.},
        title = "{DAOPHOT: A Computer Program for Crowded-Field Stellar Photometry}",
      journal = {\pasp},
         year = 1987,
        month = mar,
       volume = {99},
        pages = {191},
          doi = {10.1086/131977},
       adsurl = {https://ui.adsabs.harvard.edu/abs/1987PASP...99..191S}
}

@ARTICLE{Stewart2008,
       author = {{Stewart}, Kyle R. and {Bullock}, James S. and {Wechsler}, Risa H. and {Maller}, Ariyeh H. and {Zentner}, Andrew R.},
        title = "{Merger Histories of Galaxy Halos and Implications for Disk Survival}",
      journal = {\apj},
         year = 2008,
        month = aug,
       volume = {683},
       number = {2},
        pages = {597-610},
          doi = {10.1086/588579},
archivePrefix = {arXiv},
       eprint = {0711.5027},
 primaryClass = {astro-ph},
       adsurl = {https://ui.adsabs.harvard.edu/abs/2008ApJ...683..597S}
}

@article{Walmsley2021,
   title={Galaxy Zoo DECaLS: Detailed visual morphology measurements from volunteers and deep learning for 314 000 galaxies},
   volume={509},
   ISSN={1365-2966},
   url={http://dx.doi.org/10.1093/mnras/stab2093},
   DOI={10.1093/mnras/stab2093},
   number={3},
   journal={\mnras},
   publisher={Oxford University Press (OUP)},
   author={Walmsley, Mike and Lintott, Chris and Géron, Tobias and Kruk, Sandor and Krawczyk, Coleman and Willett, Kyle W and Bamford, Steven and Kelvin, Lee S and Fortson, Lucy and Gal, Yarin and Keel, William and Masters, Karen L and Mehta, Vihang and Simmons, Brooke D and Smethurst, Rebecca and Smith, Lewis and Baeten, Elisabeth M and Macmillan, Christine},
   year={2021},
   month=sep, pages={3966–3988}
}

@software{photutils,
  author       = {Larry Bradley},
  title        = {astropy/photutils: 1.8.0},
  month        = may,
  year         = 2023,
  publisher    = {Zenodo},
  version      = {1.8.0},
  doi          = {10.5281/zenodo.7946442},
  url          = {https://doi.org/10.5281/zenodo.7946442},
}

@article{lynden-bell_ghostly_1995,
    title = {Ghostly streams from the formation of the {Galaxy}’s halo},
    volume = {275},
    issn = {0035-8711},
    url = {https://doi.org/10.1093/mnras/275.2.429},
    doi = {10.1093/mnras/275.2.429},
    number = {2},
    journal = {\mnras},
    author = {Lynden-Bell, D. and Lynden-Bell, R. M.},
    month = jul,
    year = {1995},
    pages = {429--442},
}

@article{johnston_disruption_1995,
    title = {The {Disruption} of the {Sagittarius} {Dwarf} {Galaxy}},
    volume = {451},
    doi = {10.1086/176247},
    journal = {\apj},
    author = {Johnston, Kathryn V. and Spergel, David N. and Hernquist, Lars},
    month = oct,
    year = {1995},
    pages = {598},
}

@article{johnston_fossil_1996,
    title = {Fossil {Signatures} of {Ancient} {Accretion} {Events} in the {Halo}},
    volume = {465},
    doi = {10.1086/177418},
    journal = {\apj},
    author = {Johnston, Kathryn V. and Hernquist, Lars and Bolte, Michael},
    month = jul,
    year = {1996},
    pages = {278},
}

@article{helmi_stellar_2008,
    title = {The stellar halo of the {Galaxy}},
    volume = {15},
    doi = {10.1007/s00159-008-0009-6},
    number = {3},
    journal = {\aapr},
    author = {Helmi, Amina},
    month = jun,
    year = {2008},
    pages = {145--188},
}

@article{merrifield_measuring_1998,
    title = {Measuring galaxy potentials using shell kinematics},
    volume = {297},
    doi = {10.1046/j.1365-8711.1998.01625.x},
    number = {4},
    journal = {\mnras},
    author = {Merrifield, Michael R. and Kuijken, Konrad},
    month = jul,
    year = {1998},
    pages = {1292--1296},
}

@article{ebrova_quadruple-peaked_2012,
    title = {Quadruple-peaked spectral line profiles as a tool to constrain gravitational potential of shell galaxies},
    volume = {545},
    doi = {10.1051/0004-6361/201219940},
    journal = {\aap},
    author = {Ebrová, I. and Jílková, L. and Jungwiert, B. and Křížek, M. and Bílek, M. and Bartošková, K. and Skalická, T. and Stoklasová, I.},
    month = sep,
    year = {2012},
    pages = {A33},
}

@article{sanderson_analytical_2013,
    title = {An analytical phase-space model for tidal caustics},
    volume = {435},
    issn = {0035-8711},
    url = {https://doi.org/10.1093/mnras/stt1307},
    doi = {10.1093/mnras/stt1307},
    number = {1},
    journal = {\mnras},
    author = {Sanderson, Robyn E. and Helmi, Amina},
    month = aug,
    year = {2013},
    pages = {378--399},
}

@article{bonaca_milky_2014,
    title = {{MILKY} {WAY} {MASS} {AND} {POTENTIAL} {RECOVERY} {USING} {TIDAL} {STREAMS} {IN} {A} {REALISTIC} {HALO}},
    volume = {795},
    url = {https://dx.doi.org/10.1088/0004-637X/795/1/94},
    doi = {10.1088/0004-637X/795/1/94},
    number = {1},
    journal = {\apj},
    author = {Bonaca, Ana and Geha, Marla and Küpper, Andreas H. W. and Diemand, Jürg and Johnston, Kathryn V. and Hogg, David W.},
    month = oct,
    year = {2014},
    pages = {94},
}

@article{pearson_tidal_2015,
    title = {{TIDAL} {STREAM} {MORPHOLOGY} {AS} {AN} {INDICATOR} {OF} {DARK} {MATTER} {HALO} {GEOMETRY}: {THE} {CASE} {OF} {PALOMAR}-5},
    volume = {799},
    url = {https://dx.doi.org/10.1088/0004-637X/799/1/28},
    doi = {10.1088/0004-637X/799/1/28},
    number = {1},
    journal = {\apj},
    author = {Pearson, Sarah and Küpper, Andreas H. W. and Johnston, Kathryn V. and Price-Whelan, Adrian M.},
    month = jan,
    year = {2015},
    pages = {28},
}

@article{pearson_gaps_2017,
    title = {Gaps and length asymmetry in the stellar stream {Palomar} 5 as effects of {Galactic} bar rotation},
    volume = {1},
    doi = {10.1038/s41550-017-0220-3},
    journal = {Nature Astronomy},
    author = {Pearson, Sarah and Price-Whelan, Adrian M. and Johnston, Kathryn V.},
    month = aug,
    year = {2017},
    pages = {633--639},
}

@article{bonaca_information_2018,
    title = {The {Information} {Content} in {Cold} {Stellar} {Streams}},
    volume = {867},
    url = {https://dx.doi.org/10.3847/1538-4357/aae4da},
    doi = {10.3847/1538-4357/aae4da},
    number = {2},
    journal = {\apj},
    author = {Bonaca, Ana and Hogg, David W.},
    month = nov,
    year = {2018},
    pages = {101},
}

@article{pearson_mapping_2022,
    title = {Mapping {Dark} {Matter} with {Extragalactic} {Stellar} {Streams}: {The} {Case} of {Centaurus} {A}},
    volume = {941},
    url = {https://dx.doi.org/10.3847/1538-4357/ac9bfb},
    doi = {10.3847/1538-4357/ac9bfb},
    number = {1},
    journal = {\apj},
    author = {Pearson, Sarah and Price-Whelan, Adrian M. and Hogg, David W. and Seth, Anil C. and Sand, David J. and Hunt, Jason A. S. and Crnojević, Denija},
    month = dec,
    year = {2022},
    pages = {19},
}

@article{nibauer_constraining_2023,
    title = {Constraining the {Gravitational} {Potential} from the {Projected} {Morphology} of {Extragalactic} {Tidal} {Streams}},
    volume = {954},
    url = {https://dx.doi.org/10.3847/1538-4357/ace9bc},
    doi = {10.3847/1538-4357/ace9bc},
    number = {2},
    journal = {\apj},
    author = {Nibauer, Jacob and Bonaca, Ana and Johnston, Kathryn V.},
    month = sep,
    year = {2023},
    pages = {195},
}

@article{yavetz_stream_2023,
    title = {Stream {Fanning} and {Bifurcations}: {Observable} {Signatures} of {Resonances} in {Stellar} {Stream} {Morphology}},
    volume = {954},
    url = {https://dx.doi.org/10.3847/1538-4357/ace7b9},
    doi = {10.3847/1538-4357/ace7b9},
    number = {2},
    journal = {\apj},
    author = {Yavetz, Tomer D. and Johnston, Kathryn V. and Pearson, Sarah and Price-Whelan, Adrian M. and Hamilton, Chris},
    month = sep,
    year = {2023},
    pages = {215},
}

@article{nibauer_galactic_2025,
    title = {Galactic {Accelerations} from the {GD}-1 {Stream} {Suggest} a {Tilted} {Dark} {Matter} {Halo}},
    volume = {985},
    url = {https://dx.doi.org/10.3847/2041-8213/add0a9},
    doi = {10.3847/2041-8213/add0a9},
    number = {1},
    journal = {\apjl},
    author = {Nibauer, Jacob and Bonaca, Ana},
    month = may,
    year = {2025},
    pages = {L22},
}

@article{de_jong_kilo-degree_2013,
    title = {The {Kilo}-{Degree} {Survey}},
    volume = {35},
    issn = {0922-6435},
    url = {https://ui.adsabs.harvard.edu/abs/2013ExA....35...25D},
    doi = {10.1007/s10686-012-9306-1},
    urldate = {2025-09-15},
    journal = {Exp. Astron.},
    author = {de Jong, Jelte T. A. and Verdoes Kleijn, Gijs A. and Kuijken, Konrad H. and Valentijn, Edwin A.},
    month = jan,
    year = {2013},
    pages = {25--44},
}

@article{aihara_hyper_2018,
    title = {The {Hyper} {Suprime}-{Cam} {SSP} {Survey}: {Overview} and survey design},
    volume = {70},
    issn = {0004-6264},
    shorttitle = {The {Hyper} {Suprime}-{Cam} {SSP} {Survey}},
    url = {https://ui.adsabs.harvard.edu/abs/2018PASJ...70S...4A},
    doi = {10.1093/pasj/psx066},
    urldate = {2025-09-15},
    journal = {\pasj},
    author = {Aihara, Hiroaki and Arimoto, Nobuo and Armstrong, Robert and Arnouts, Stéphane and Bahcall, Neta A. and Bickerton, Steven and Bosch, James and Bundy, Kevin and Capak, Peter L. and Chan, James H. H. and Chiba, Masashi and Coupon, Jean and Egami, Eiichi and Enoki, Motohiro and Finet, Francois and Fujimori, Hiroki and Fujimoto, Seiji and Furusawa, Hisanori and Furusawa, Junko and Goto, Tomotsugu and Goulding, Andy and Greco, Johnny P. and Greene, Jenny E. and Gunn, James E. and Hamana, Takashi and Harikane, Yuichi and Hashimoto, Yasuhiro and Hattori, Takashi and Hayashi, Masao and Hayashi, Yusuke and Hełminiak, Krzysztof G. and Higuchi, Ryo and Hikage, Chiaki and Ho, Paul T. P. and Hsieh, Bau-Ching and Huang, Kuiyun and Huang, Song and Ikeda, Hiroyuki and Imanishi, Masatoshi and Inoue, Akio K. and Iwasawa, Kazushi and Iwata, Ikuru and Jaelani, Anton T. and Jian, Hung-Yu and Kamata, Yukiko and Karoji, Hiroshi and Kashikawa, Nobunari and Katayama, Nobuhiko and Kawanomoto, Satoshi and Kayo, Issha and Koda, Jin and Koike, Michitaro and Kojima, Takashi and Komiyama, Yutaka and Konno, Akira and Koshida, Shintaro and Koyama, Yusei and Kusakabe, Haruka and Leauthaud, Alexie and Lee, Chien-Hsiu and Lin, Lihwai and Lin, Yen-Ting and Lupton, Robert H. and Mandelbaum, Rachel and Matsuoka, Yoshiki and Medezinski, Elinor and Mineo, Sogo and Miyama, Shoken and Miyatake, Hironao and Miyazaki, Satoshi and Momose, Rieko and More, Anupreeta and More, Surhud and Moritani, Yuki and Moriya, Takashi J. and Morokuma, Tomoki and Mukae, Shiro and Murata, Ryoma and Murayama, Hitoshi and Nagao, Tohru and Nakata, Fumiaki and Niida, Mana and Niikura, Hiroko and Nishizawa, Atsushi J. and Obuchi, Yoshiyuki and Oguri, Masamune and Oishi, Yukie and Okabe, Nobuhiro and Okamoto, Sakurako and Okura, Yuki and Ono, Yoshiaki and Onodera, Masato and Onoue, Masafusa and Osato, Ken and Ouchi, Masami and Price, Paul A. and Pyo, Tae-Soo and Sako, Masao and Sawicki, Marcin and Shibuya, Takatoshi and Shimasaku, Kazuhiro and Shimono, Atsushi and Shirasaki, Masato and Silverman, John D. and Simet, Melanie and Speagle, Joshua and Spergel, David N. and Strauss, Michael A. and Sugahara, Yuma and Sugiyama, Naoshi and Suto, Yasushi and Suyu, Sherry H. and Suzuki, Nao and Tait, Philip J. and Takada, Masahiro and Takata, Tadafumi and Tamura, Naoyuki and Tanaka, Manobu M. and Tanaka, Masaomi and Tanaka, Masayuki and Tanaka, Yoko and Terai, Tsuyoshi and Terashima, Yuichi and Toba, Yoshiki and Tominaga, Nozomu and Toshikawa, Jun and Turner, Edwin L. and Uchida, Tomohisa and Uchiyama, Hisakazu and Umetsu, Keiichi and Uraguchi, Fumihiro and Urata, Yuji and Usuda, Tomonori and Utsumi, Yousuke and Wang, Shiang-Yu and Wang, Wei-Hao and Wong, Kenneth C. and Yabe, Kiyoto and Yamada, Yoshihiko and Yamanoi, Hitomi and Yasuda, Naoki and Yeh, Sherry and Yonehara, Atsunori and Yuma, Suraphong},
    month = jan,
    year = {2018},
    pages = {S4},
}

@article{dark_energy_survey_collaboration_dark_2016,
    title = {The {Dark} {Energy} {Survey}: more than dark energy - an overview},
    volume = {460},
    issn = {0035-8711},
    shorttitle = {The {Dark} {Energy} {Survey}},
    url = {https://ui.adsabs.harvard.edu/abs/2016MNRAS.460.1270D},
    doi = {10.1093/mnras/stw641},
    urldate = {2025-09-15},
    journal = {\mnras},
    author = {{Dark Energy Survey Collaboration} and Abbott, T. and Abdalla, F. B. and Aleksić, J. and Allam, S. and Amara, A. and Bacon, D. and Balbinot, E. and Banerji, M. and Bechtol, K. and Benoit-Lévy, A. and Bernstein, G. M. and Bertin, E. and Blazek, J. and Bonnett, C. and Bridle, S. and Brooks, D. and Brunner, R. J. and Buckley-Geer, E. and Burke, D. L. and Caminha, G. B. and Capozzi, D. and Carlsen, J. and Carnero-Rosell, A. and Carollo, M. and Carrasco-Kind, M. and Carretero, J. and Castander, F. J. and Clerkin, L. and Collett, T. and Conselice, C. and Crocce, M. and Cunha, C. E. and D'Andrea, C. B. and da Costa, L. N. and Davis, T. M. and Desai, S. and Diehl, H. T. and Dietrich, J. P. and Dodelson, S. and Doel, P. and Drlica-Wagner, A. and Estrada, J. and Etherington, J. and Evrard, A. E. and Fabbri, J. and Finley, D. A. and Flaugher, B. and Foley, R. J. and Fosalba, P. and Frieman, J. and García-Bellido, J. and Gaztanaga, E. and Gerdes, D. W. and Giannantonio, T. and Goldstein, D. A. and Gruen, D. and Gruendl, R. A. and Guarnieri, P. and Gutierrez, G. and Hartley, W. and Honscheid, K. and Jain, B. and James, D. J. and Jeltema, T. and Jouvel, S. and Kessler, R. and King, A. and Kirk, D. and Kron, R. and Kuehn, K. and Kuropatkin, N. and Lahav, O. and Li, T. S. and Lima, M. and Lin, H. and Maia, M. A. G. and Makler, M. and Manera, M. and Maraston, C. and Marshall, J. L. and Martini, P. and McMahon, R. G. and Melchior, P. and Merson, A. and Miller, C. J. and Miquel, R. and Mohr, J. J. and Morice-Atkinson, X. and Naidoo, K. and Neilsen, E. and Nichol, R. C. and Nord, B. and Ogando, R. and Ostrovski, F. and Palmese, A. and Papadopoulos, A. and Peiris, H. V. and Peoples, J. and Percival, W. J. and Plazas, A. A. and Reed, S. L. and Refregier, A. and Romer, A. K. and Roodman, A. and Ross, A. and Rozo, E. and Rykoff, E. S. and Sadeh, I. and Sako, M. and Sánchez, C. and Sanchez, E. and Santiago, B. and Scarpine, V. and Schubnell, M. and Sevilla-Noarbe, I. and Sheldon, E. and Smith, M. and Smith, R. C. and Soares-Santos, M. and Sobreira, F. and Soumagnac, M. and Suchyta, E. and Sullivan, M. and Swanson, M. and Tarle, G. and Thaler, J. and Thomas, D. and Thomas, R. C. and Tucker, D. and Vieira, J. D. and Vikram, V. and Walker, A. R. and Wechsler, R. H. and Weller, J. and Wester, W. and Whiteway, L. and Wilcox, H. and Yanny, B. and Zhang, Y. and Zuntz, J.},
    month = aug,
    year = {2016},
    pages = {1270--1299},
}

@article{martinez-delgado_hidden_2023,
    title = {Hidden depths in the local {Universe}: {The} {Stellar} {Stream} {Legacy} {Survey}},
    volume = {671},
    issn = {0004-6361},
    shorttitle = {Hidden depths in the local {Universe}},
    url = {https://ui.adsabs.harvard.edu/abs/2023A&A...671A.141M},
    doi = {10.1051/0004-6361/202245011},
    urldate = {2025-09-15},
    journal = {\aap},
    author = {Martínez-Delgado, David and Cooper, Andrew P. and Román, Javier and Pillepich, Annalisa and Erkal, Denis and Pearson, Sarah and Moustakas, John and Laporte, Chervin F. P. and Laine, Seppo and Akhlaghi, Mohammad and Lang, Dustin and Makarov, Dmitry and Borlaff, Alejandro S. and Donatiello, Giuseppe and Pearson, William J. and Miró-Carretero, Juan and Cuillandre, Jean-Charles and Domínguez, Helena and Roca-Fàbrega, Santi and Frenk, Carlos S. and Schmidt, Judy and Gómez-Flechoso, María A. and Guzman, Rafael and Libeskind, Noam I. and Dey, Arjun and Weaver, Benjamin A. and Schlegel, David and Myers, Adam D. and Valdes, Frank G.},
    month = mar,
    year = {2023},
    pages = {A141},
}

@article{hayat_self-supervised_2021,
    title = {Self-supervised {Representation} {Learning} for {Astronomical} {Images}},
    volume = {911},
    doi = {10.3847/2041-8213/abf2c7},
    number = {2},
    journal = {\apjl},
    author = {Hayat, Md Abul and Stein, George and Harrington, Peter and Lukić, Zarija and Mustafa, Mustafa},
    month = apr,
    year = {2021},
    pages = {L33},
}

@article{willett_galaxy_2017,
    title = {Galaxy {Zoo}: morphological classifications for 120 000 galaxies in {HST} legacy imaging★},
    volume = {464},
    issn = {0035-8711},
    shorttitle = {Galaxy {Zoo}},
    url = {https://doi.org/10.1093/mnras/stw2568},
    doi = {10.1093/mnras/stw2568},
    number = {4},
    urldate = {2025-10-16},
    journal = {\mnras},
    author = {Willett, Kyle W. and Galloway, Melanie A. and Bamford, Steven P. and Lintott, Chris J. and Masters, Karen L. and Scarlata, Claudia and Simmons, B. D. and Beck, Melanie and Cardamone, Carolin N. and Cheung, Edmond and Edmondson, Edward M. and Fortson, Lucy F. and Griffith, Roger L. and Häußler, Boris and Han, Anna and Hart, Ross and Melvin, Thomas and Parrish, Michael and Schawinski, Kevin and Smethurst, R. J. and Smith, Arfon M.},
    month = feb,
    year = {2017},
    pages = {4176--4203},
}

@article{narayan_assessing_2021,
    title = {Assessing single-cell transcriptomic variability through density-preserving data visualization},
    volume = {39},
    copyright = {2021 The Author(s), under exclusive licence to Springer Nature America, Inc.},
    issn = {1546-1696},
    url = {https://www.nature.com/articles/s41587-020-00801-7},
    doi = {10.1038/s41587-020-00801-7},
    language = {en},
    number = {6},
    urldate = {2025-10-22},
    journal = {Nature Biotechnology},
    author = {Narayan, Ashwin and Berger, Bonnie and Cho, Hyunghoon},
    month = jun,
    year = {2021},
    pages = {765--774},
}

@ARTICLE{Ivezic2019,
       author = {{Ivezi{\'c}}, {\v{Z}}eljko and {Kahn}, Steven M. and {Tyson}, J. Anthony and {Abel}, Bob and {Acosta}, Emily and {Allsman}, Robyn and {Alonso}, David and {AlSayyad}, Yusra and {Anderson}, Scott F. and {Andrew}, John and {Angel}, James Roger P. and {Angeli}, George Z. and {Ansari}, Reza and {Antilogus}, Pierre and {Araujo}, Constanza and {Armstrong}, Robert and {Arndt}, Kirk T. and {Astier}, Pierre and {Aubourg}, {\'E}ric and {Auza}, Nicole and {Axelrod}, Tim S. and {Bard}, Deborah J. and {Barr}, Jeff D. and {Barrau}, Aurelian and {Bartlett}, James G. and {Bauer}, Amanda E. and {Bauman}, Brian J. and {Baumont}, Sylvain and {Bechtol}, Ellen and {Bechtol}, Keith and {Becker}, Andrew C. and {Becla}, Jacek and {Beldica}, Cristina and {Bellavia}, Steve and {Bianco}, Federica B. and {Biswas}, Rahul and {Blanc}, Guillaume and {Blazek}, Jonathan and {Blandford}, Roger D. and {Bloom}, Josh S. and {Bogart}, Joanne and {Bond}, Tim W. and {Booth}, Michael T. and {Borgland}, Anders W. and {Borne}, Kirk and {Bosch}, James F. and {Boutigny}, Dominique and {Brackett}, Craig A. and {Bradshaw}, Andrew and {Brandt}, William Nielsen and {Brown}, Michael E. and {Bullock}, James S. and {Burchat}, Patricia and {Burke}, David L. and {Cagnoli}, Gianpietro and {Calabrese}, Daniel and {Callahan}, Shawn and {Callen}, Alice L. and {Carlin}, Jeffrey L. and {Carlson}, Erin L. and {Chandrasekharan}, Srinivasan and {Charles-Emerson}, Glenaver and {Chesley}, Steve and {Cheu}, Elliott C. and {Chiang}, Hsin-Fang and {Chiang}, James and {Chirino}, Carol and {Chow}, Derek and {Ciardi}, David R. and {Claver}, Charles F. and {Cohen-Tanugi}, Johann and {Cockrum}, Joseph J. and {Coles}, Rebecca and {Connolly}, Andrew J. and {Cook}, Kem H. and {Cooray}, Asantha and {Covey}, Kevin R. and {Cribbs}, Chris and {Cui}, Wei and {Cutri}, Roc and {Daly}, Philip N. and {Daniel}, Scott F. and {Daruich}, Felipe and {Daubard}, Guillaume and {Daues}, Greg and {Dawson}, William and {Delgado}, Francisco and {Dellapenna}, Alfred and {de Peyster}, Robert and {de Val-Borro}, Miguel and {Digel}, Seth W. and {Doherty}, Peter and {Dubois}, Richard and {Dubois-Felsmann}, Gregory P. and {Durech}, Josef and {Economou}, Frossie and {Eifler}, Tim and {Eracleous}, Michael and {Emmons}, Benjamin L. and {Fausti Neto}, Angelo and {Ferguson}, Henry and {Figueroa}, Enrique and {Fisher-Levine}, Merlin and {Focke}, Warren and {Foss}, Michael D. and {Frank}, James and {Freemon}, Michael D. and {Gangler}, Emmanuel and {Gawiser}, Eric and {Geary}, John C. and {Gee}, Perry and {Geha}, Marla and {Gessner}, Charles J.~B. and {Gibson}, Robert R. and {Gilmore}, D. Kirk and {Glanzman}, Thomas and {Glick}, William and {Goldina}, Tatiana and {Goldstein}, Daniel A. and {Goodenow}, Iain and {Graham}, Melissa L. and {Gressler}, William J. and {Gris}, Philippe and {Guy}, Leanne P. and {Guyonnet}, Augustin and {Haller}, Gunther and {Harris}, Ron and {Hascall}, Patrick A. and {Haupt}, Justine and {Hernandez}, Fabio and {Herrmann}, Sven and {Hileman}, Edward and {Hoblitt}, Joshua and {Hodgson}, John A. and {Hogan}, Craig and {Howard}, James D. and {Huang}, Dajun and {Huffer}, Michael E. and {Ingraham}, Patrick and {Innes}, Walter R. and {Jacoby}, Suzanne H. and {Jain}, Bhuvnesh and {Jammes}, Fabrice and {Jee}, M. James and {Jenness}, Tim and {Jernigan}, Garrett and {Jevremovi{\'c}}, Darko and {Johns}, Kenneth and {Johnson}, Anthony S. and {Johnson}, Margaret W.~G. and {Jones}, R. Lynne and {Juramy-Gilles}, Claire and {Juri{\'c}}, Mario and {Kalirai}, Jason S. and {Kallivayalil}, Nitya J. and {Kalmbach}, Bryce and {Kantor}, Jeffrey P. and {Karst}, Pierre and {Kasliwal}, Mansi M. and {Kelly}, Heather and {Kessler}, Richard and {Kinnison}, Veronica and {Kirkby}, David and {Knox}, Lloyd and {Kotov}, Ivan V. and {Krabbendam}, Victor L. and {Krughoff}, K. Simon and {Kub{\'a}nek}, Petr and {Kuczewski}, John and {Kulkarni}, Shri and {Ku}, John and {Kurita}, Nadine R. and {Lage}, Craig S. and {Lambert}, Ron and {Lange}, Travis and {Langton}, J. Brian and {Le Guillou}, Laurent and {Levine}, Deborah and {Liang}, Ming and {Lim}, Kian-Tat and {Lintott}, Chris J. and {Long}, Kevin E. and {Lopez}, Margaux and {Lotz}, Paul J. and {Lupton}, Robert H. and {Lust}, Nate B. and {MacArthur}, Lauren A. and {Mahabal}, Ashish and {Mandelbaum}, Rachel and {Markiewicz}, Thomas W. and {Marsh}, Darren S. and {Marshall}, Philip J. and {Marshall}, Stuart and {May}, Morgan and {McKercher}, Robert and {McQueen}, Michelle and {Meyers}, Joshua and {Migliore}, Myriam and {Miller}, Michelle and {Mills}, David J.},
        title = "{LSST: From Science Drivers to Reference Design and Anticipated Data Products}",
      journal = {\apj},
         year = 2019,
        month = mar,
       volume = {873},
       number = {2},
          eid = {111},
        pages = {111},
          doi = {10.3847/1538-4357/ab042c},
archivePrefix = {arXiv},
       eprint = {0805.2366},
 primaryClass = {astro-ph},
       adsurl = {https://ui.adsabs.harvard.edu/abs/2019ApJ...873..111I}
}

@article{gwyn_unions_2025,
    title = {{UNIONS}: {The} {Ultraviolet} {Near}-infrared {Optical} {Northern} {Survey}},
    volume = {170},
    url = {https://doi.org/10.3847/1538-3881/ae03ab},
    doi = {10.3847/1538-3881/ae03ab},
    number = {6},
    journal = {\aj},
    author = {Gwyn, Stephen and McConnachie, Alan W. and Cuillandre, Jean-Charles and Chambers, Kenneth C. and Magnier, Eugene A. and de Boer, Thomas and Hudson, Michael J. and Oguri, Masamune and Furusawa, Hisanori and Hildebrandt, Hendrik and Carlberg, Raymond and Ellison, Sara L. and Furusawa, Junko and Gavazzi, Raphaël and Ibata, Rodrigo and Mellier, Yannick and Osato, Ken and Aussel, H. and Baumont, Lucie and Bayer, Manuel and Boulade, Olivier and Côté, Patrick and Chemaly, David and Daley, Cail and Duc, Pierre-Alain and Durret, Florence and Ellien, A. and Fabbro, Sébastien and Ferreira, Leonardo and Fitriana, Itsna K. and Le Floc’h, Emeric and Fudamoto, Yoshinobu and Gao, Hua and Goh, L. W. K. and Goto, Tomotsugu and Guerrini, Sacha and Guinot, Axel and Hénault-Brunet, Vincent and Hammer, Francois and Harikane, Yuichi and Hayashi, Kohei and Heesters, Nick and Ichikawa, Kohei and Kilbinger, Martin and Kuzma, P. B. and Li, Qinxun and Liaudat, Tobías I. and Lin, Chien-Cheng and Müller, Oliver and Martin, Nicolas F. and Matsuoka, Yoshiki and Medina, Gustavo E. and Miyatake, Hironao and Miyazaki, Satoshi and Mpetha, Charlie T. and Nagao, Tohru and Navarro, Julio F. and Niwano, Masafumi and Ogami, Itsuki and Okabe, Nobuhiro and Onoue, Masafusa and Paek, Gregory S. H. and Parker, Laura C. and Patton, David R. and Peters, Fabian Hervas and Prunet, Simon and Sánchez-Janssen, Rubén and Schultheis, M. and Sestito, Federico and Smith, Simon E. T. and Starck, J.-L. and Starkenburg, Else and Stone, Connor and Storfer, Christopher and Suzuki, Yoshihisa and Erben, T. and Taibi, Salvatore and Thomas, G. F. and Toba, Yoshiki and Uchiyama, Hisakazu and Valls-Gabaud, David and Venn, Kim A. and Van Waerbeke, Ludovic and Wainscoat, Richard J. and Wilkinson, Scott and Wittje, Anna and Yoshida, Taketo and Zhang, TianFang and Zhong, Yuxing},
    month = nov,
    year = {2025},
    pages = {324},
}

@article{barnes_transformations_1992,
    title = {Transformations of galaxies. {I} - {Mergers} of equal-mass stellar disks},
    volume = {393},
    issn = {0004-637X, 1538-4357},
    url = {http://adsabs.harvard.edu/doi/10.1086/171522},
    doi = {10.1086/171522},
    language = {en},
    urldate = {2026-01-06},
    journal = {\apj},
    author = {Barnes, Joshua E.},
    month = jul,
    year = {1992},
    pages = {484},
}

@article{barnes_encounters_1988,
    title = {Encounters of disk/halo galaxies},
    volume = {331},
    issn = {0004-637X, 1538-4357},
    url = {http://adsabs.harvard.edu/doi/10.1086/166593},
    doi = {10.1086/166593},
    language = {en},
    urldate = {2026-01-06},
    journal = {\apj},
    author = {Barnes, Joshua E.},
    month = aug,
    year = {1988},
    pages = {699},
}

@inproceedings{fitzpatrick_noao_2014,
    title = {The {NOAO} {Data} {Laboratory}: a conceptual overview},
    volume = {9149},
    booktitle = {Observatory Operations: Strategies, Processes, and Systems V},
    shorttitle = {The {NOAO} {Data} {Laboratory}},
    url = {https://ui.adsabs.harvard.edu/abs/2014SPIE.9149E..1TF},
    doi = {10.1117/12.2057445},
    urldate = {2026-01-21},
    author = {Fitzpatrick, Michael J. and Olsen, Knut and Economou, Frossie and Stobie, Elizabeth B. and Beers, T. C. and Dickinson, Mark and Norris, Patrick and Saha, Abi and Seaman, Robert and Silva, David R. and Swaters, Robert A. and Thomas, Brian and Valdes, Francisco},
    month = aug,
    year = {2014},
    pages = {91491T},
}

@article{nikutta_data_2020,
    title = {Data {Lab}—{A} community science platform},
    volume = {33},
    issn = {2213-1337},
    url = {https://www.sciencedirect.com/science/article/pii/S2213133720300652},
    doi = {10.1016/j.ascom.2020.100411},
    urldate = {2026-01-21},
    journal = {A\&C},
    author = {Nikutta, R. and Fitzpatrick, M. and Scott, A. and Weaver, B. A.},
    month = oct,
    year = {2020},
    pages = {100411},
}

@article{shannon_mathematical_1948,
    title = {A mathematical theory of communication},
    volume = {27},
    url = {https://ui.adsabs.harvard.edu/abs/1948BSTJ...27..623S},
    doi = {10.1002/j.1538-7305.1948.tb00917.x},
    urldate = {2026-01-29},
    journal = {Bell Labs Technical Journal},
    author = {Shannon, C. E.},
    month = oct,
    year = {1948},
    pages = {623--656},
}

@article{walmsley_identification_2019,
    title = {Identification of low surface brightness tidal features in galaxies using convolutional neural networks},
    volume = {483},
    doi = {10.1093/mnras/sty3232},
    number = {3},
    journal = {\mnras},
    author = {Walmsley, Mike and Ferguson, Annette M. N. and Mann, Robert G. and Lintott, Chris J.},
    month = mar,
    year = {2019},
    pages = {2968--2982},
}

@article{gwyn_canadafrancehawaii_2012,
    title = {{THE} {CANADA}–{FRANCE}–{HAWAII} ℡{ESCOPE} {LEGACY} {SURVEY}: {STACKED} {IMAGES} {AND} {CATALOGS}},
    volume = {143},
    issn = {1538-3881},
    shorttitle = {{THE} {CANADA}–{FRANCE}–{HAWAII} ℡{ESCOPE} {LEGACY} {SURVEY}},
    url = {https://doi.org/10.1088/0004-6256/143/2/38},
    doi = {10.1088/0004-6256/143/2/38},
    language = {en},
    number = {2},
    urldate = {2026-04-21},
    journal = {\aj},
    author = {Gwyn, Stephen D. J.},
    month = jan,
    year = {2012},
    pages = {38},
}

@inproceedings{chen_intriguing_2021,
 author = {Chen, Ting and Luo, Calvin and Li, Lala},
 booktitle = {Advances in Neural Information Processing Systems},
 pages = {11834--11845},
 title = {Intriguing Properties of Contrastive Losses},
 url = {https://proceedings.neurips.cc/paper_files/paper/2021/file/628f16b29939d1b060af49f66ae0f7f8-Paper.pdf},
 volume = {34},
 year = {2021}
}

\begin{appendix}
\onecolumn
\section{Calibrating the tiers} \label{appendix:calibrating_the_tiers}

\begin{table*}[h!]
    \centering
    \caption{\label{tab:A1:sigmoid_host_galaxies}Host galaxies used for calibrating the tiered sigmoid scaling function across $g$-, $r$-, and $i$-band in each tier.}
    \begin{tabular*}{\textwidth}{@{\extracolsep{\fill}} l c c c @{}}
        \hline\hline
        Tier & $g$-band & $r$-band & $i$-band \\
        \hline
        1 & PGC010807 & PGC008973 & PGC066522 \\
          & PGC009634 & PGC597851 & PGC128532 \\
          & PGC017993 & PGC006030 & PGC009063 \\
        \hline
        2 & PGC013094 & PGC067194 & PGC127531 \\
          & PGC069579 & PGC127531 & PGC015980 \\
          & PGC128520 & PGC015980 & PGC015602 \\
        \hline
        3 & PGC131085 & PGC064979 & PGC064979 \\
          & PGC128506 & PGC132026 & PGC128506 \\
          & PGC132026 & PGC131331 & PGC135102 \\
        \hline
        4 & PGC064979 & PGC452979 & PGC452979 \\
          & PGC452979 & -         & -         \\
        \hline\hline
    \end{tabular*}
    \tablefoot{The fourth tier contains fewer objects per bin due to lack of samples from \cite{Miro-Carretero2024} which have a host galaxy with a magnitude within the bounds of the fourth tier.}
\end{table*}

\begin{table*}[h!]
    \centering
    \caption{\label{tab:A1:sigmoid_statistics}Mean medians and mean median absolute deviations of $g$-, $r$-, and $i$-band measurements across four tiers, in picomaggies.}
    \begin{tabular*}{\textwidth}{@{\extracolsep{\fill}} l c c c c c c c c c @{}}
        \hline
        \hline
        Tier & $\langle$Median$\rangle_g$ & $\langle$Median$\rangle_r$ & $\langle$Median$\rangle_i$ & $\langle$MAD$\rangle_g$ & $\langle$MAD$\rangle_r$ & $\langle$MAD$\rangle_i$ & $\langle$S/N$\rangle_g$ & $\langle$S/N$\rangle_r$ & $\langle$S/N$\rangle_i$ \\
        \hline
        1 & 0.6563 & 2.1420 & 2.0170 & 0.2903 & 0.9901 & 0.8790 & 4740.8 & 4791.9 & 2562.0 \\
        2 & 0.5567 & 3.6980 & 1.7670 & 0.3897 & 1.6865 & 1.2669 & 2664.5 & 2387.2 & 1759.6 \\
        3 & 0.8530 & 0.8989 & 1.3952 & 0.4055 & 0.5064 & 0.8723 & 1716.5 & 1753.2 & 1245.5 \\
        4 & 0.8572 & 1.8735 & 1.4057 & 0.4565 & 0.8063 & 0.9384 & 826.97 & 905.67 & 515.79 \\
        \hline
    \end{tabular*}
    \tablefoot{The mean medians and mean median absolute deviations are obtained by averaging photometric statistics of elliptical apertures around stellar streams. The objects belonging to each tier are outlined in Table \ref{tab:A1:sigmoid_host_galaxies}. The mean S/N for each band is obtained by averaging the reported \texttt{snr\_\{g,r,i\}} values in the \texttt{des\_dr2.main} table \citep{DES_DR2} for each object used for calibration in its respective tier.}
\end{table*}

\section{Tier randomization ablation} \label{appendix:tier_randomization_ablation}
\begin{figure}[h!]
    \centering
    \begin{tabular}{cc}
        \includegraphics[width=0.44\textwidth]{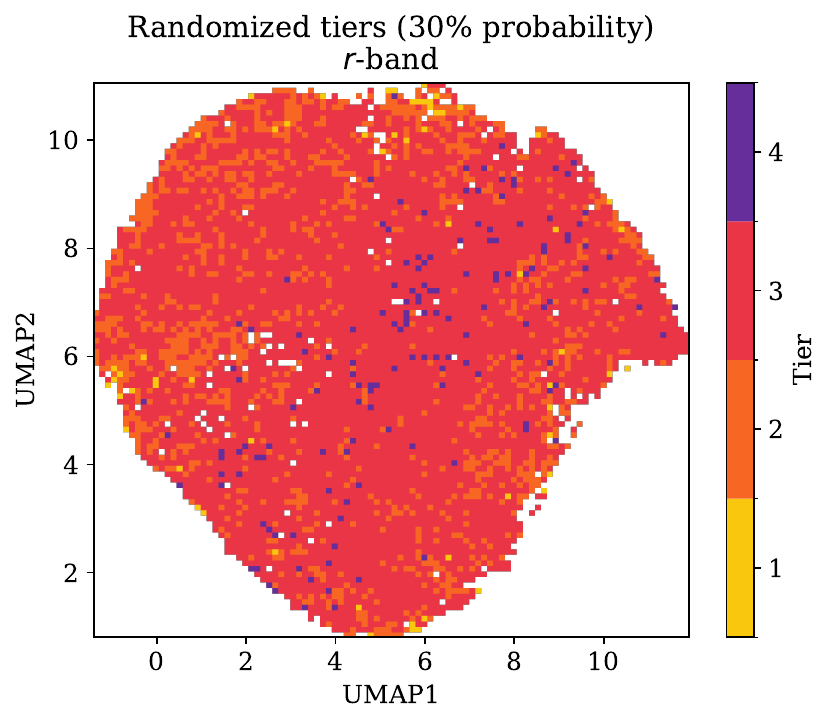} & 
        \includegraphics[width=0.45\textwidth]{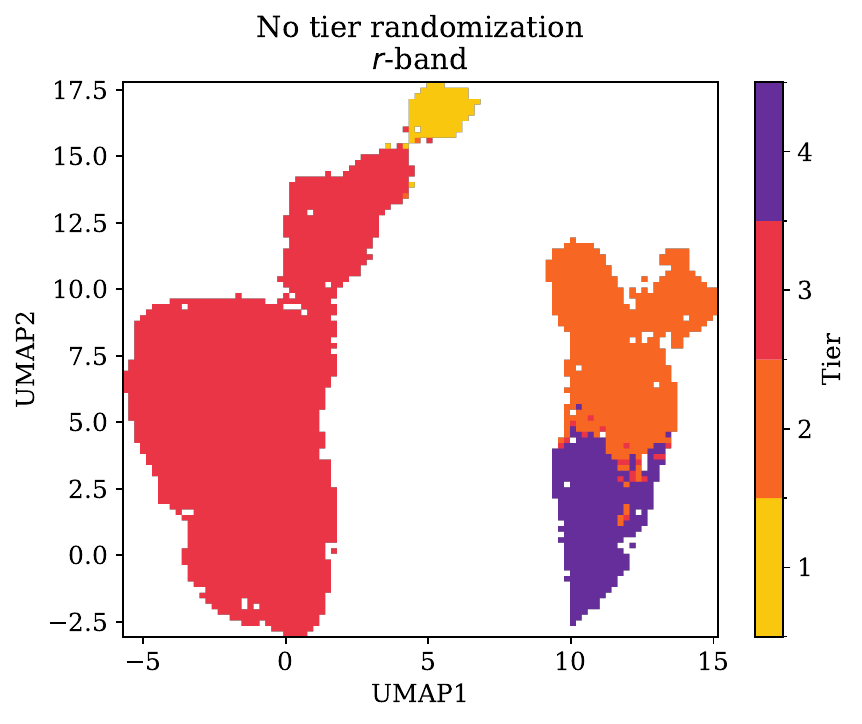}
    \end{tabular}
    \caption{(Left) UMAP projection of the embeddings from the NNCLR encoder trained with tier randomization enabled shows significantly less clustering than (right) when the NNCLR encoder is trained with no tier randomization. As the tiers are meant to improve the representation of low surface brightness features without needing to individually tune the scaling function to each image in the dataset, we do not want the model to focus on the categorization of the tiers. Tier randomization is therefore an important hyperparameter when applying the tiered sigmoid scaling function with a contrastive learning framework.}
    \label{fig:3:tier_ablation_histograms}
\end{figure}
As described in Section \ref{subsec:image_preprocessing}, the tiered sigmoid scaling function introduces a spurious category into the data: the tiers. A significant challenge in the detection of extragalactic stellar streams (both with machine learning and through visual inspection) is the need to manually finetune a scaling function for each image, such that low-surface-brightness features become apparent. The tiers of the tiered sigmoid scaling function are designed to remove the need to finetune on a per-image basis. As we do not aim for the NNCLR encoder to focus on discriminating different tiers, we assigned a 30\% probability of randomly selecting a tier for any given augmented image. We chose a low probability since we wished to discourage tier memorization, but not completely remove the information that the different tiers may provide in finding low-surface-brightness features.  In this section, we demonstrate the difference in the resulting UMAP projections when the tier randomization is removed.

Figure \ref{fig:3:tier_ablation_histograms} demonstrates the resulting UMAP projection of the latent representations formed by a model trained without tier randomization. We have binned the UMAP embeddings into a median histogram of the $r$-band tier; however, the results are similar across bands. As described in Section \ref{subsec:nnclr}, we initialized the same model parameters across models and used the same global seed for the training of the NNCLR encoder and the UMAP projection for direct comparison. It is immediately apparent that the model trained without tier randomization has clustered samples with similar tiers. 

\section{Single-image saliency maps}\label{appendix:single_image_saliency_maps}
\begin{figure}[h!]
    \centering
    \begin{tabular}{c}
        \includegraphics[width=\textwidth]{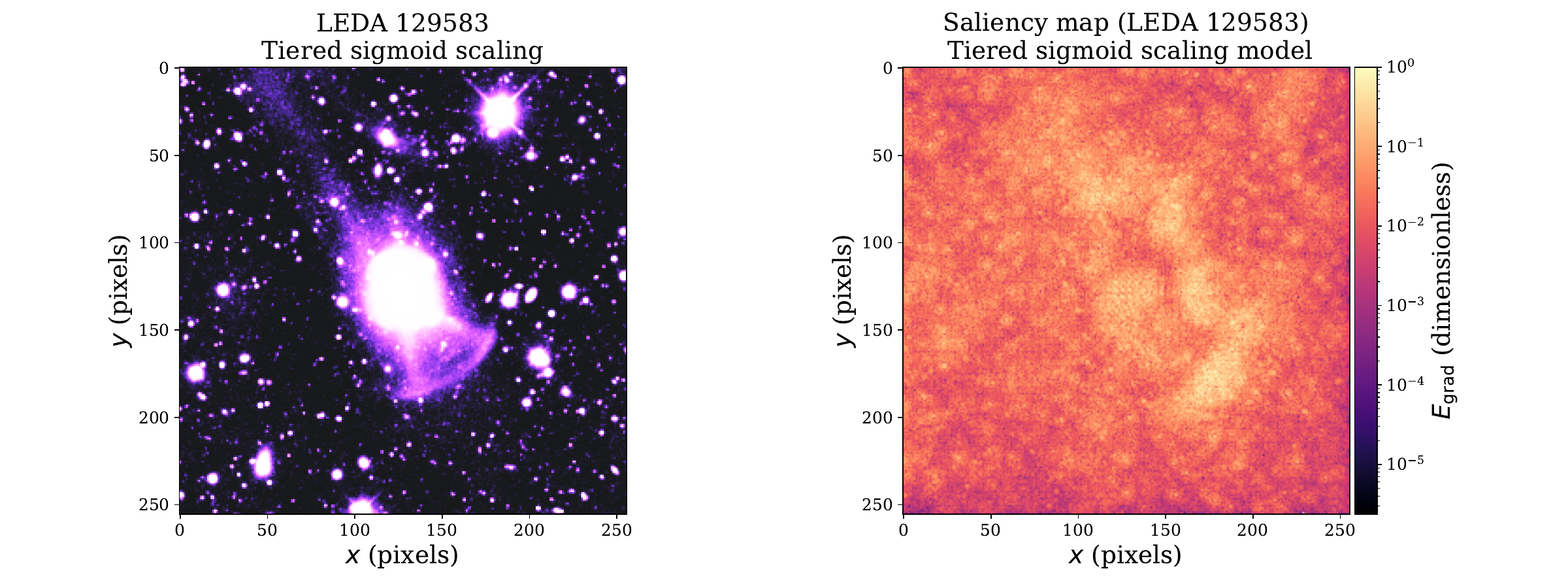} \\
        \includegraphics[width=\textwidth]{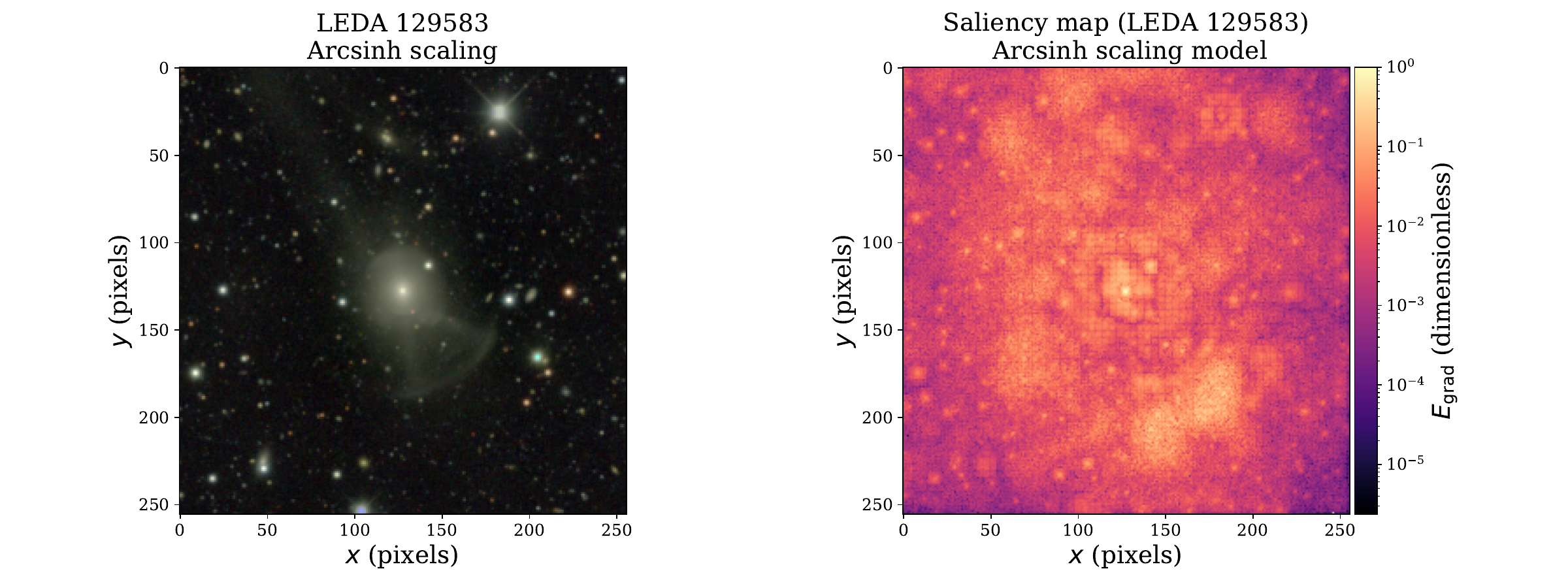}
    \end{tabular}
    \caption{Saliency maps of the (top row) tiered sigmoid scaling model and (bottom row) arcsinh scaling model for the image cutout of LEDA 129583. The collimated tidal features at the bottom-right and top-left of the galaxy are difficult to distinguish in both of the saliency maps, since the latent representation vectors from each NNCLR encoder encapsulate many features of the image.}
    \label{fig:saliency_map_single_image_TSS}
\end{figure}
\twocolumn
\end{appendix}
\end{document}